\documentclass[preprint,12pt]{elsarticle}
\usepackage{diagbox} 
\usepackage{lineno,hyperref}
\usepackage{bm}
\usepackage{amsfonts}
\usepackage{amssymb}
\usepackage{amsmath}
\usepackage{isomath}

\usepackage{scalerel,stackengine}
\stackMath
\newcommand\reallywidehat[1]{%
\savestack{\tmpbox}{\stretchto{%
  \scaleto{%
    \scalerel*[\widthof{\ensuremath{#1}}]{\kern-.6pt\bigwedge\kern-.6pt}%
    {\rule[-\textheight/2]{1ex}{\textheight}}%WIDTH-LIMITED BIG WEDGE
  }{\textheight}% 
}{0.5ex}}%
\stackon[1pt]{#1}{\tmpbox}%
}

\usepackage{stmaryrd} 
\usepackage[usenames]{color}
\usepackage{epsfig}
\usepackage{graphicx}
\usepackage{psfrag}

\graphicspath{{Figures/}}
\usepackage{float}
\floatstyle{boxed}
\newfloat{algorithm}{htb}{loa}
\floatname{algorithm}{Algorithm}
\usepackage{algorithm,algpseudocode}
\usepackage{caption}
\usepackage{subcaption}
\usepackage{color}
\usepackage{xcolor}
\usepackage{multirow}
\usepackage{natbib}
\modulolinenumbers[5]

\usepackage[left = 2cm, right=2.5cm, top=2cm, bottom =2cm]{geometry}

\begin{document}
\begin{frontmatter}

\title{Mixed formulation of physics-informed neural networks for thermo-mechanically coupled systems and heterogeneous domains} 

% A mixed formulation for physics-informed neural networks to solve a multi-physics problem with employing a sequential training

\author{ Ali Harandi$^{1,*}$, Ahmad Moeineddin$^{2}$, Michael Kaliske$^{2}$, Stefanie Reese$^1$, Shahed Rezaei$^{3,*}$}
\address{$^1$Institute of Applied Mechanics, \\ RWTH Aachen University, Mies-van-der-Rohe-Str. 1, D-52074 Aachen, Germany}
\address{$^2$Institute for Structural Analysis, Technical University of Dresden, \\Georg-Schumann-Str. 7, D-01187 Dresden, Germany}
\address{$^3$Access e.V, Intze-Str. 5, 52072 Aachen, Germany}
\address{$^*$ corresponding authors: ali.harandi@ifam.rwth-aachen.de, s.rezaei@access-technology.de}
%\corref{ali.harandi@ifam.rwth-aachen.de, Shahed.rezaeitu-darmstadt.de}

\begin{abstract}
Physics-informed neural networks (PINNs) are a new tool for solving boundary value problems by defining loss functions of neural networks based on governing equations, boundary conditions, and initial conditions. Recent investigations have shown that when designing loss functions for many engineering problems, using first-order derivatives and combining equations from both strong and weak forms can lead to much better accuracy, especially when there are heterogeneity and variable jumps in the domain. This new approach is called the mixed formulation for PINNs, which takes ideas from the mixed finite element method. In this method, the PDE is reformulated as a system of equations where the primary unknowns are the fluxes or gradients of the solution, and the secondary unknowns are the solution itself.
In this work, we propose applying the mixed formulation to solve multi-physical problems, specifically a stationary thermo-mechanically coupled system of equations. Additionally, we discuss both sequential and fully coupled unsupervised training and compare their accuracy and computational cost. To improve the accuracy of the network, we incorporate hard boundary constraints to ensure valid predictions. We then investigate how different optimizers and architectures affect accuracy and efficiency.
Finally, we introduce a simple approach for parametric learning that is similar to transfer learning. This approach combines data and physics to address the limitations of PINNs regarding computational cost and improves the network's ability to predict the response of the system for unseen cases. The outcomes of this work will be useful for many other engineering applications where deep learning is employed on multiple coupled systems of equations for fast and reliable computations.
\end{abstract} 
\begin{keyword} 
Physics-informed neural networks, Thermo-mechanically coupled problems, Heterogeneous solids, Hard constraints, Parametric learning
\end{keyword}

\end{frontmatter}

%\newpage
%%%%%%%%%%%%%%%%%%%%%%%%%%%%%%%%%%%%%%%%%%%%%%%%
\section{Introduction} 
% intro to the topic: the importance of DL
Deep learning (DL) methods are a branch of machine learning (ML) that have remarkable flexibility in approximating arbitrary functions and operators \cite{Anima2020, lu2021learning}. They are able to perform the latter task by finding the hidden patterns for the training data set (also known as supervised learning). Moreover, DL methods are capable of discovering the solution to a boundary value problem (BVP) by solely considering physical constraints (unsupervised learning). Therefore, DL models find their way into many engineering applications from material design to structural analysis \cite{Bock2019}. One advantage of ML algorithms compared to the classical solver is the huge speedup that one achieves after successful training. The efficiency of these models makes them extremely relevant for multi-scaling analysis where the bottleneck is in the process of information transformation between the upper scale and the lower scale (see \cite{fernandez2020, Peng2021, mianroodi2022lossless} and references therein). An important issue is that the network has to be trained for enough observations considering all the relevant scenarios. Otherwise, one cannot completely rely on the network outcome. Adding well-known physical constraints to the data training process improves the network's prediction capabilities (see \cite{REZAEI2022PINN} and references therein). The latter point needs to be investigated properly for different test cases and physical problems to gain the most advantage from applying the physical laws to the classical data-driven approaches. 

Here, we intend to examine the potential of the DL method for the field of computational engineering, where we deal with a complex heterogeneous domain as well as a coupled system of governing equations. In what follows, we shall review some of the major contributions and address the new contributions of the current study.

% DATA-DRIVEN
Training ML algorithms based on the available physical data obtained from our numerical solvers can speed up the process of predicting the solution. \citet{PARK2022111192} utilized a multi-scale kernel neural network to enhance the prediction of strain fields for unseen cases. In traditional ML approaches, where NNs are only trained based on data, the underlying physics of the problem is embedded in the data itself, which is the result of other numerical solvers.
\citet{Yang21} employed the DL method to predict complex stress and strain fields in composites. \citet{khorrami2022artificial} 
employed a convolutional neural network to predict the stress fields for history-dependent material behavior. See also \cite{RASHID2022}, where authors employed 
a Fourier neural operator to predict a stress field at a high resolution by performing the training based on data of low resolution. NN surrogate models are powerful models to solve complex systems of equations in coupled thermal, chemical, and hydrological engineering problems \cite{Lu_2021}. Data-driven models are used to estimate the state of charge based on the driving conditions \cite{RAGONE2021} or as digital twins of complex systems like proton exchange membrane fuel cells \cite{WANG2020}.
See also \cite{LU2022114778, Kapoor2022} for a comparison between the performance of different neural operators which are suitable for industrial applications.

% PHYSICS-DRIVEN and possible enhancements
Interestingly enough, following the idea of physics-informed neural networks (PINNs) \cite{RAISSI2019}, and by employing the physical laws in the final loss functions, one can reasonably rely on the network outcome. This is due to the fact that the network outcomes automatically respect and satisfy important physical constraints. However, a proper and in-depth theoretical understanding of the convergence and generalization properties of the PINNs method is still under investigation. Therefore, the idea behind PINNs is also further extended and explored by several authors to improve the network performance even for unseen scenarios. For some examples see \cite{Ameya2020, JAGTAP2020, HAGHIGHAT2021}. In case the underlying physics of the problem is completely known and complete, the NN can be trained without any initial data and solely based on the given BVP \cite{REZAEI2022PINN}. The latter point is also referred to as unsupervised learning since no initial solution to the problem is needed beforehand. 
\citet{Hu2022} initiate discussions on how extended PINNs that work based on domain decomposition method show effectiveness in modeling multi-scale and multi-physical problems. The authors introduce a tradeoff for generalization. On the one hand, the decomposition of a complex PDE solution into several simple parts decreases the complexity and on the other hand, decomposition might lead to overfitting and the obtained solution may become less generalizable, see also investigations by \citet{JAGTAP2020, Jagtap201, Henkes2022} and \citet{WANG2022115491} related to the idea of domain decomposition.
\citet{WANG21} investigated how PINNs are biased towards learning functions along the dominant eigen-directions of the neural tangent kernel and proposed architectures that can lead to robust and accurate PINNs models. 
Different loss terms in the PINNs formulation may be treated differently on their importance.
\citet{McClenny22} proposed to utilize adaptive weights for each training point, so the neural network learns which regions of the solution are difficult and is forced to focus on them. The basic idea here is to make the weights increase as the corresponding losses increase. \citet{XIANG21} also addressed that the PINNs' performance is affected by the weighting of losses and established Gaussian probabilistic models to define the self-adaptive loss functions through the adaptive weights for each loss term. See also \cite{Bischof21, Rohrhofer21}. Another promising extension is the idea of reducing the differential orders for the construction of loss functions. For this purpose, researchers combined different ideas from the variational formulation of partial differential equations and apply them to various problems in engineering applications \cite{SAMANIEGO2020112790}. The idea is also further extended to a so-called mixed formulation, where both strong form and weak form are utilized simultaneously \cite{REZAEI2022PINN,  Henkes2022,FUHG2022110839,Abueidda2023}.

% Multiphysics problems and PINNs
One important aspect of engineering applications is to consider the multi-physical characteristics. In other words, in many realistic problems, various coupled and highly nonlinear fields have to be considered simultaneously. As an example in solid mechanics, besides the mechanical deformation, often one needs to take into account the influence of the thermal field \cite{Ruan2022}, chemical concentrations \cite{REZAEI2021104612}, electrical and/or magnetic fields \cite{Guo2022}, damage field \cite{GOSWAMI2022114587, REZAEI2022108177}, and others. 

The idea of employing PINNs for solving underlying equations within the multi-physical context is addressed by several authors. In such problems, we are dealing with different loss terms which represent different physics and might be completely different from each other in terms of the magnitude of numerical values. Therefore, a balance between different loss terms is often necessary. Therefore, classical PINNs models need to be further improved.
\citet{Amini2022} investigated the application of PINNs to find the solution to problems involving thermo–hydro-mechanical processes. The authors proposed to use a dimensionless version of the governing equations as well as a sequential training strategy to obtain better accuracy in the multiobjective optimization problem. \citet{Raj22} studied the thermo-mechanical problem using PINNs for functionally graded material and the ability of the network to approximate the displacements and showed that the network leads to an order higher error when it comes to approximating the stress fields.
\citet{Laubscher21} presented a single and segregated network for PINNs dry air humidification fluid properties heat diffusion to predict momentum, species, and temperature distributions of a dry air humidification problem. It is reported that the segregated network has lower losses when compared to the single network. \citet{MATTEY2022} also reported that the PINN’s accuracy suffers for strongly non-linear and higher-order PDEs such as Allen Cahn and Cahn Hilliard equations. The authors proposed to solve the PDE sequentially over successive time segments using a single neural network. See also studied by \citet{CHEN2022116918}, \citet{Nguyen2022} and \citet{Bischof21}. In modeling crack propagation in solids via a smeared approach, one usually assigns a field variable to damage. In this case, equations for the evolution of the damage field are usually coupled to those from the mechanical equilibrium \cite{REZAEI2022108177}. \citet{GOSWAMI2022114587} proposed a physics-informed variational formulation of DeepONet for brittle fracture analysis, see also investigations by \cite{ZHENG2022107282, Ghaffari2022}.

%%%%%% TL
The computational efficiency and cost of the training can be improved by utilizing a transfer learning algorithm \cite{Zhuang21}.
The key idea of transfer learning is to capture the common features and behavior of the source domain and transfer it to a new target domain. To do so, one can transfer the pre-trained hyperparameters of the NN and initialize them as hyperparameters of the NN for the new domain. The latter decreases the computational cost of training \cite{REZAEI2022PINN, Neyshabur20}. \citet{Tang2022} utilized transfer learning-PINNs to solve vortex-induced vibration problems.
\citet{GOSWAMI20TPF} employed the idea of transfer learning to partially retrain the network's hyperparameters to identify the crack patterns in a domain. \citet{Pellegrin2022} exploited the transfer learning idea to solve a system of equations by having various configurations (initial conditions) and showed that utilizing the trained multi-head network leads to a significant speed-up for branched flow simulations without losing accuracy.
\citet{Desai21} utilized transfer learning to solve linear systems of PDEs such as Poisson and Schrodinger equations by PINNs. \citet{Gao22} proposed a singular-valued-based transfer learning algorithm for PINNs and showed its performance for Allen Cahn equations. 

\begin{figure}[H] 
  \centering
  \includegraphics[width=0.99\linewidth]{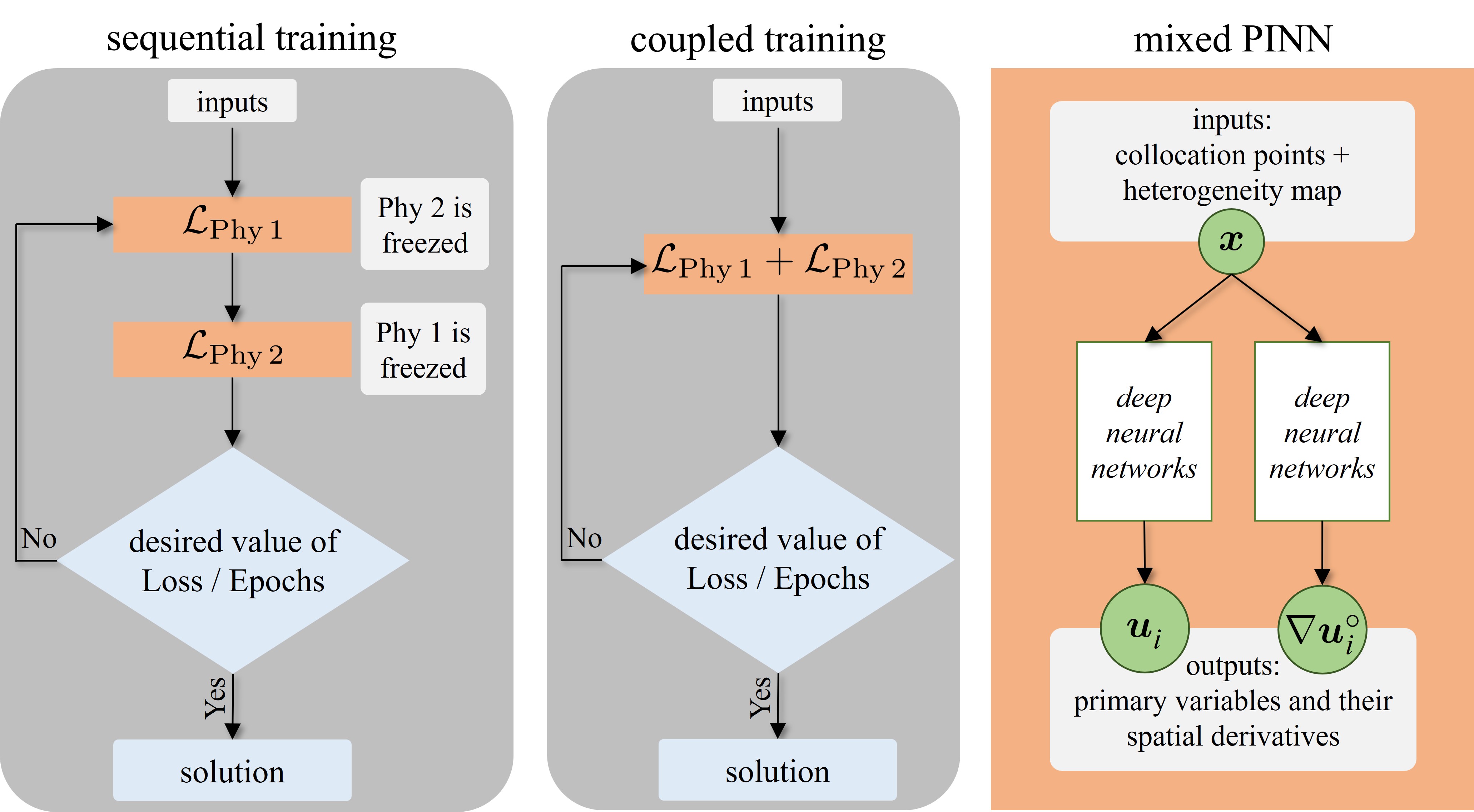}
  \caption{Sequential training and coupled training algorithms for solving a multi-physical boundary value problem as well as the mixed PINNs architecture. $\mathcal{L}_{\text{Phy}}$ stands for the relevant physics law. The variable $\bm{x}$ denotes the location of collocation points. The variable $\bm{u}_i$ shows the $i$-th primary variable (solution) and $\nabla \bm{u}_i^o$ is their spatial gradient. Separated neural networks are then utilized to predict primary variables and their spatial gradients. }
  \label{fig:intro}
\end{figure}

% overview of the paper
According to the reviewed articles, it becomes clear that having fast and reliable results by employing the PINNs method for engineering applications has to be done with care. The naive version of PINNs may not converge to the correct solution, and even so, the training process may take significantly more time compared to other numerical solvers based on the finite element method. This work discusses ideas on how to overcome the previously mentioned issues. Here, we intend to focus on the idea of first-order PINNs formulation for a stationary thermo-mechanically coupled system in heterogeneous solids and compare the performance of the coupled and sequential training (see also Fig.~\ref{fig:intro}). In Section 2, we provide the derivation of the governing equations and their variational form. In Section 3, the architecture of the new mixed formulation of PINNs is described. The results and conclusions are then provided in Sections 4 and 5, respectively.

%\citet{HAGHIGHATAmini22}
%\citet{Bischof2021} evaluated different methods for balancing the contributions of multiple terms of the PINNs loss function and proposed some techniques for loss balancing of PINNs. The authors concluded that proper loss balancing resulted in faster training and higher accuracy.
%\citet{Giampaolo2022} %Grey-Scott system solution is a reaction-diffusion problem in which two evolution equations exist
%\citet{HE2020} % the DL two data then minimize a realation which is based on data again read their abstract again
%\citet{Haber2022} %this is for CNN to a coupled scalar transport equation with Navier stokes 
%\citet{WAHEED21} %localy adaptive weighting and activations to enhance the convergence rate and accuracy of pinns for eikonal equation
%\citet{Rohrhofer21} % study theeffectiveness of self adaptive weighting and activation functions   
%\citet{JAGTAP20} % they have employed the adaptive activation functions 
% More details on the construction of the loss functions are provided throughout the paper.
%%%%%%%%%%%%%%%%%%%%%%%%%%%%%%%%%%%%%%%%%
\section{Formulation of the problem}
\subsection{Stationary thermo-elasticity in heterogeneous solids}
The formulation behind the thermo-mechanical problem is first summarized in what follows. In this problem, we are dealing with the displacement field denoted by the vector $\bm{u}$ and the temperature field denoted by the scalar parameter $T$. Through different coupling terms, these two fields influence each other. Starting with kinematics, the total strain $\bm{\varepsilon}$ is additively decomposed into the elastic part and the thermal part as
\begin{align}
\label{kinematics}
\bm{\varepsilon}\,=\,\bm{\varepsilon}_e\,+\,\bm{\varepsilon}_t =\,\text{sym}\left(\text{grad}(\bm{u})\right)\,=\,\cfrac{1}{2}\left(\nabla \bm{u}\,+\,\nabla \bm{u}^T\right).
\end{align} 
In Eq.\,(\ref{kinematics}), $\bm{\varepsilon}_e$ represents the elastic strain tensor, and $\bm{\varepsilon}_t$ presents the thermal strain caused by the temperature field. Considering isotropic material behavior, the thermal strain tensor follows a linear expansion law and reads 
\begin{align}
\label{eps_t}
\bm{\varepsilon}_t\,=\,\alpha(x,y)\left(T-T_0\right)\,\bm{I}.
\end{align} 
Here, $\alpha$ is the thermal expansion coefficient and it can vary spatially as we are dealing with a heterogeneous microstructure. Moreover, $T_0$ is the initial temperature and $\boldsymbol I$ is the second-order identity tensor.
Utilizing the fourth-order elasticity tensor $\mathbb{C}$,
one can write the stress tensor as
\begin{align}
\label{materiallaw4}  
\bm{\sigma}\,=\mathbb{C}(x,y)\,\left(\bm{\varepsilon}-\bm{\varepsilon}_t\right).
\end{align}
Note, that at this point, we assume the elastic properties to be temperature-independent. Considering temperature-dependent material properties shall be investigated in future work and the investigations in the current work can easily be extended to such problems. Writing Eq.\,(\ref{materiallaw4}) in Voigt notation reads
\begin{align}
\label{materiallaw}
\hat{\bm{\sigma}}\,=\hat{{C}}(x,y)~\left(\hat{\bm{\varepsilon}}\,-\,\hat{\bm{\varepsilon}}_t\right).
\end{align} 
In Eq.\,(\ref{materiallaw}), $\hat{\bullet}$ presents the tensor ($\bullet$) in the Voigt notation. For the two-dimensional plane strain setup, the position-dependent elasticity tensor $\hat{\mathcal{C}}$ is written as
\begin{align}
\label{elasticityPlanestrain}
\hat{{C}}(x,y)\,=\, \dfrac{E(x,y)}{(1-2\nu(x,y))(1+\nu(x,y))}\,\begin{bmatrix} 1-\nu(x,y) & \nu(x,y) & 0 \\
\nu(x,y) & 1-\nu(x,y) & 0 \\
0 & 0 & \dfrac{1-2\nu(x,y)}{2}
\end{bmatrix}.
\end{align}
Here, Young's modulus $E$ and Poisson’s ratio $\nu$ represent the elastic constants of the material. The isotropic assumption holds for every point of material.
It is worth mentioning that having a heterogeneous domain causes the dependency of material parameters on the coordinates of collocation points. For the mechanical field, the balance of linear momentum by having no body force vector, and the Dirichlet and the Neumann boundary conditions read
\begin{align}  
\label{Equilbrium} 
\text{div}({\bm{\sigma}})\ &= \bm{0}~~~~\text{in}~~~ \Omega,\\
\label{BcsMech_d}
\bm{u} &= \bar{\bm{u}}~~~\text{on}~~\Gamma_D, \\ 
\label{BcsMech_n}
\bm{\sigma} \cdot \bm{n} = \bm{t} &= \bar{\bm{t}}~~~~\text{on}~~\Gamma_N.
\end{align} 

Next, we write the equation of energy balance in the absence of a heat source, and for a steady-state problem (\cite{Ruan2022})
\begin{align}
\label{StrongfromThermal}
\text{div}(\bm{q}) &= 0~~~~ \text{in}~ \Omega, \\
\label{BCsthermalD}
T &= \bar{T}~~~\text{on}~ \bar{\Gamma}_D, \\
\label{BCsthermalN}
\bm{q}\cdot \bm{n} = q_n &= \bar{q}~~~~\text{on}~ \bar{\Gamma}_N.
\end{align} 

In the above equations, $\Omega$, and $\Gamma$ are the material points in the body and on the boundary of the mechanical field, respectively. The temperature boundary condition is also satisfied on $\bar{\Gamma}$. In Eqs.\,(\ref{StrongfromThermal})\, and\,(\ref{BCsthermalN}), $\bm{q}$ denotes the heat flux and is defined based on Fourier's law as 
\begin{align}
\label{Fourier}
\bm{q} = -k(x,y)\,\nabla T.
\end{align} 
Here, $k(x,y)$ shows the position-dependent heat conductivity coefficient.

To obtain the Galerkin weak form (WF) for Eq.\,(\ref{Equilbrium}), the multiplication with a test function $\delta\bm{u}$ is done, which also satisfies the Dirichlet boundary conditions of the mechanical field. After integration by parts and employing the Neumann boundary conditions, one finally gets
\begin{align}
\label{Weakform}
\int_{\Omega} \delta{\bm{\hat{\varepsilon}}}^T_e\,\hat{{{C}}}(x,y)\,\hat{\bm{\varepsilon}_e}\,~dV\,-\,\int_{\Gamma_N} \delta{\bm{u}^T}\,\bar{\bm{t}}~dA=0. 
\end{align} 
Considering the balance between the mechanical internal energy $E^M_{\text{int}}$ and the mechanical external energy $E^M_{\text{ext}}$, we write the weak form in Eq.\,(\ref{Weakform}) as a minimization of the total energy, see \cite{SAMANIEGO2020112790}, as
\begin{align}
\label{int_Energy}
\text{Minimize:~}&\underbrace{\int_{\Omega}\dfrac{1}{2}\hat{\bm{\varepsilon}_e}^T\,\hat{{{C}}}(x,y)\,\hat{\bm{\varepsilon}_e}\,~dV}_{E^M_{\text{int}}} -
\underbrace{\int_{\Gamma_N} {\bm{u}^T}\,\bar{\bm{t}}~dA}_{E^M_{\text{ext}}},
\end{align} 
which is subjected to BCs in Eq.\,(\ref{BcsMech_d}) and Eq.\,(\ref{BcsMech_n}).
By multiplying the Eq.\,(\ref{StrongfromThermal}) with a test function $\delta T$ and following a similar procedure for the mechanical field and applying the boundary condition in Eqs.\,(\ref{BCsthermalD})\, and\,\ref{BCsthermalN}, the Galerkin weak form for the thermal field leads to 
\begin{align}
\label{weakformthermal}
\int_{\Omega}\,k(x,y)\,\nabla^T T\,\delta(\nabla T)~dV\,+\,\int_{\bar{\Gamma}_N}\bar{q}~\delta T~dA\,=\,0.
\end{align}
One can also reformulate the weak form in Eq.\,(\ref{weakformthermal}) as a variation of the energy. The latter initiates the balance of thermal internal and external energies, which leads to
\begin{align}
\label{int_Energy_ther}
\text{Minimize:~}&\underbrace{\int_{\Omega}\dfrac{1}{2}\,k(x,y)\,\nabla^T T\,\nabla T\,~dV}_{E^T_{\text{int}}} + 
\underbrace{\int_{\bar{\Gamma}_N} \bar{q}\,T~dA}_{E^T_{\text{ext}}}.
\end{align}
The above equation is minimized subjected to BCs in Eq.\,(\ref{BCsthermalD}) and Eq.\,(\ref{BCsthermalN}).

In Eq.\,(\ref{int_Energy}), the minimization is performed for both the displacement field $\bm{u}$ and temperature field $T$ in the coupled training process. However, by using sequential training and fixing one field variable, the referring loss function is minimized with respect to the displacement field $\bm{u}$. For Eq.\,(\ref{weakformthermal}), in the absence of the other fields in the formulation, the minimization is done with respect to the temperature field $T$ in both training procedures.

Similarly to the previous works \cite{REZAEI2022PINN,SAMANIEGO2020112790}, expressions in 
Eqs.\,(\ref{Weakform})\, and\,(\ref{weakformthermal}) are added to the neural network's loss function for the primary variables $\bm{u}$ and $T$, respectively. 

\begin{figure}[H] 
  \centering
  \includegraphics[width=0.75\linewidth]{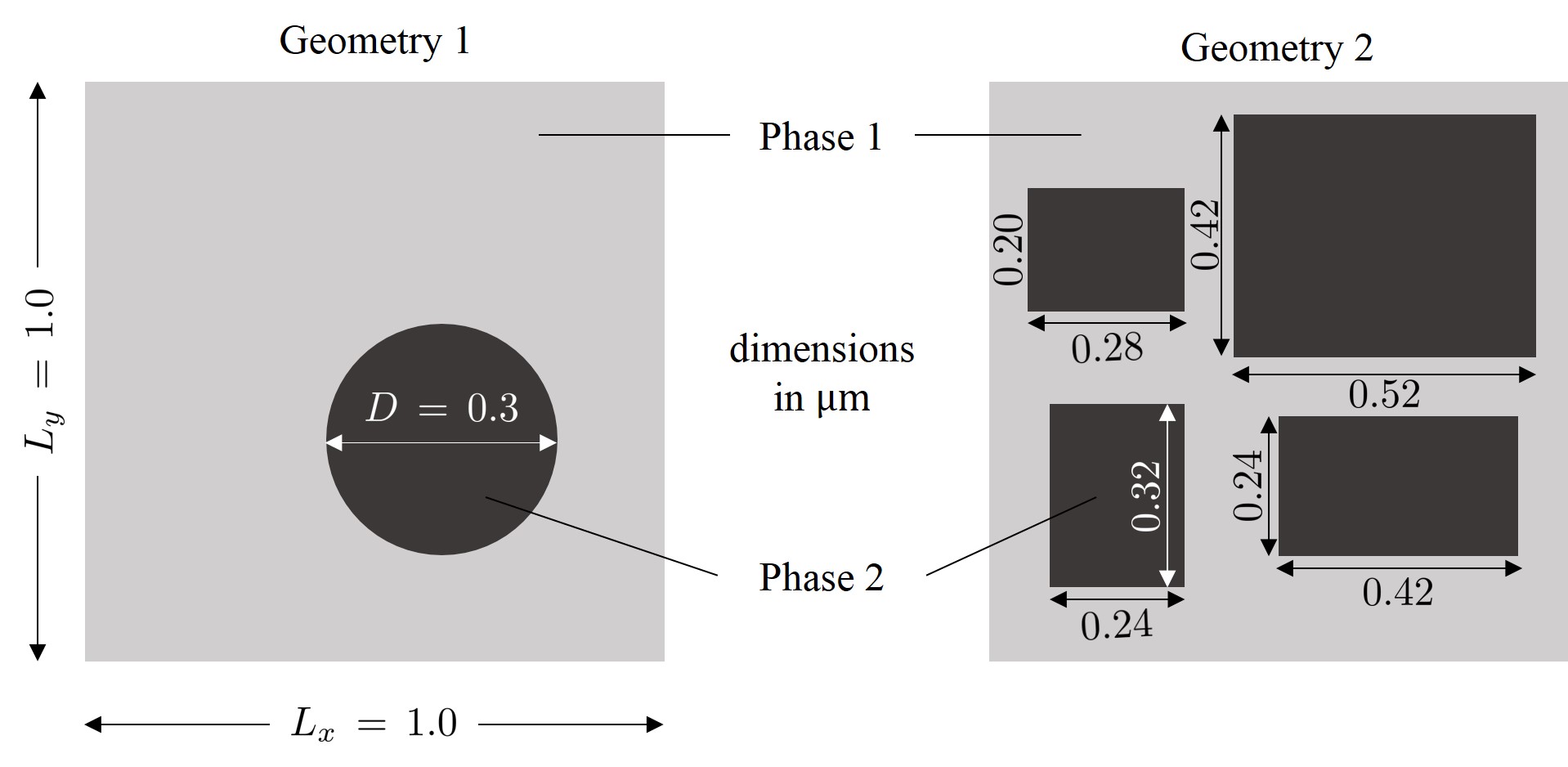}
  \caption{Selected geometries for studies on the thermal elasticity problem. The color variation shows the different values of Young's modulus and thermal conductivity.}
  \label{fig:allgeom}
\end{figure}

\begin{figure}[H] 
  \centering
  \includegraphics[width=0.75\linewidth]{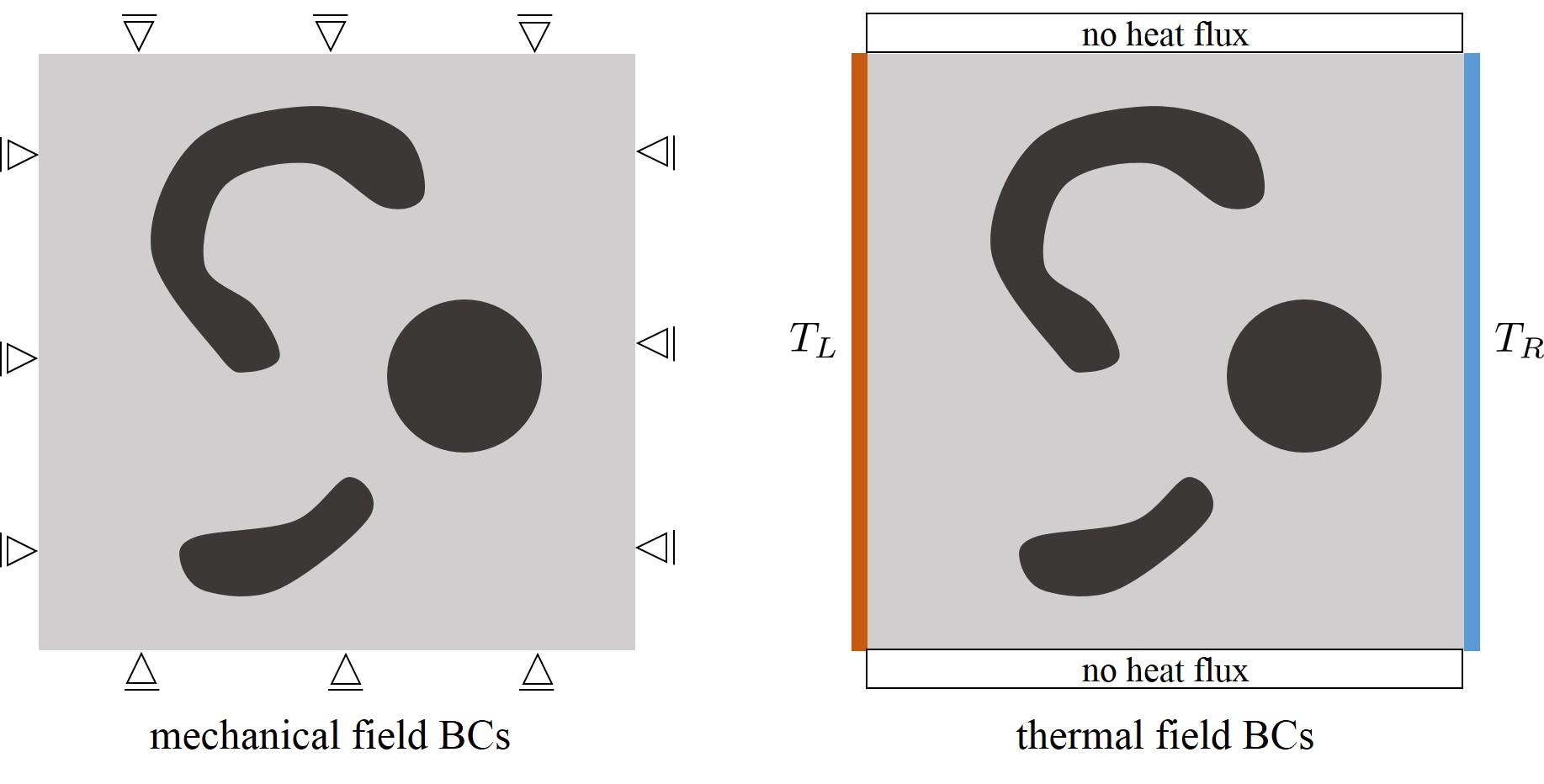}
  \caption{Boundary conditions for the thermo-mechanically coupled problem. Left: mechanical field boundary conditions. Right: Thermal field boundary conditions.}
  \label{fig:BC}
\end{figure}

%%%%%%%%%%%%%%%%%%%%%%%%%%%%%%%%%%%%%%%%%%%%%%%%%%
%% Case studies %%
Two different sets of geometries according to Fig.~\ref{fig:allgeom} are considered. The first geometry is selected to represent the fiber matrix setup. The second geometry stands for an advanced engineering alloy microstructure that contains rectangular inclusions. 
Properties of phase 1 (matrix) are denoted by the sub-index "mat" while the material properties for phase 2 (inclusion) are denoted by the sub-index "inc". Next, we define the ratio $E_{\text{mat}}/E_{\text{inc}} = k_{\text{mat}}/k_{\text{inc}} = R$. The value $R$ is equal to $0.3$ and $2$ for the left and right geometries, respectively.
The stationary thermal elasticity problem is solved for these two setups. Finally, the boundary conditions are summarized in Fig.~\ref{fig:BC}. For the mechanical field, the left and right parts are restricted to move in the $x$-direction and the upper and lower edges are fixed along $y$-direction. For the thermal field, constant temperatures are applied on the left and the right edges while on the upper and lower edges, there is no heat flux normal to the surface.

%\newpage
%%%%%%%%%%%%%%%%%%%%%%%%%%%%%%%%%%%%%%%%
\section{PINNs' architecture}
\subsection{Thermo-elasticity problem} \label{sec:thermos}
\begin{figure}[H] 
  \centering
  \includegraphics[width=0.80\linewidth]{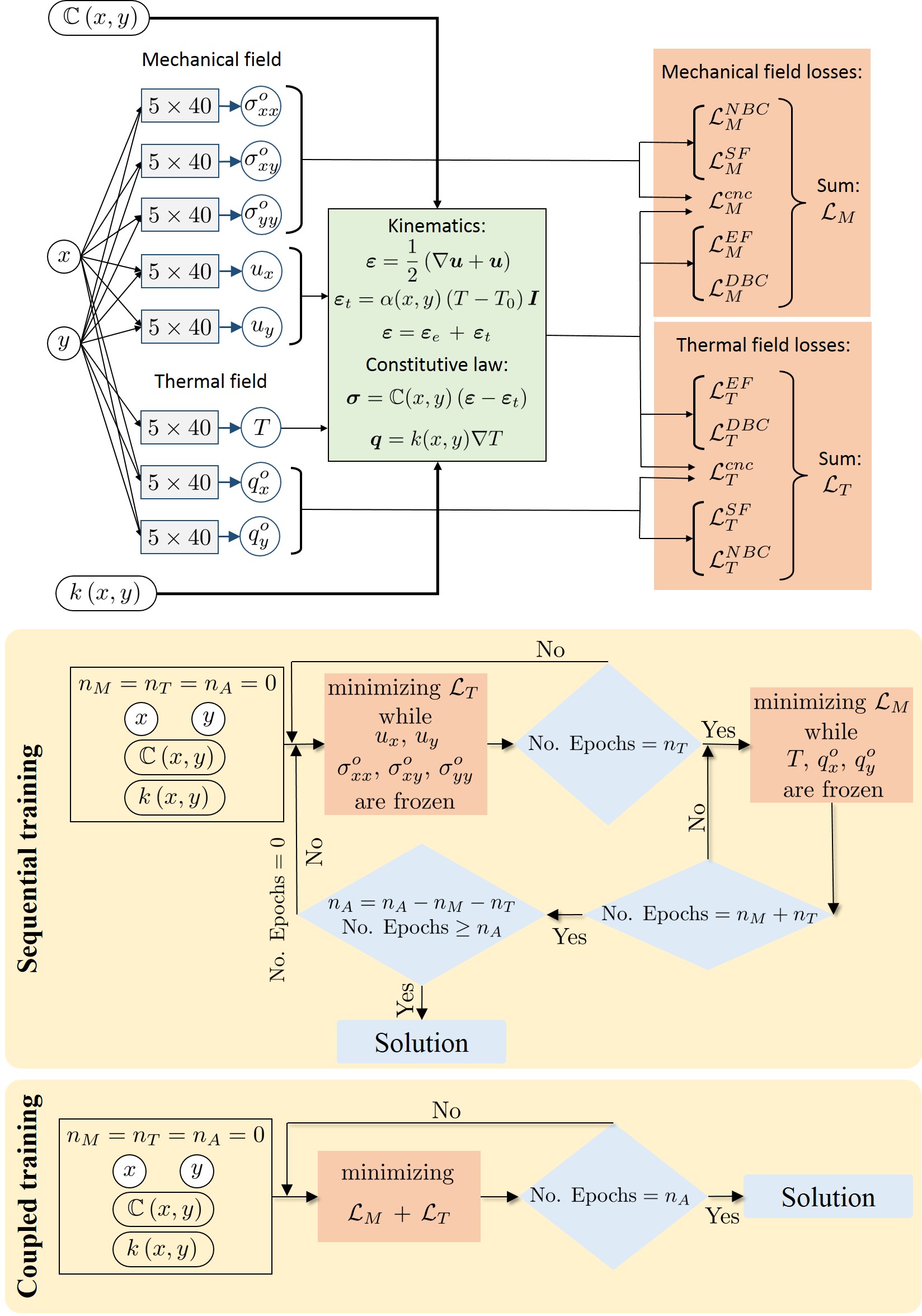}
  \caption{ Network architecture and loss functions for the multi-physical PINNs, sequential and coupled training flowcharts for the thermoelasticity problem. }
  %\vspace{-4mm}
  \label{fig:PINNs_net_ML}
\end{figure}

Based on the mixed formulation proposed by \citet{REZAEI2022PINN}, the input for the network is the location of the collocation points (i.e. the coordinates $x$ and $y$ in a 2D setting).
The outputs are then divided into the field variables from the mechanical and thermal sub-problem. For the mechanical field, outputs are the components of the displacement vector as well as the stress tensor (i.e. $u_x$, $u_y$, $\sigma_{xx}$, $\sigma_{xy}$, and $\sigma_{yy}$). For the thermal field, the output layer includes the scalar temperature value as well as the components of the heat flux (i.e. $T$, $q_x$, and $q_y$). It is important to mention that we intend to use the separate fully connected feed-forward neural network (FFNN) for each individual output variable (see also Fig.~\ref{fig:PINNs_net_ML} and explanations in \cite{REZAEI2022PINN}). The structure of each neural network takes the standard form where it can be split into a single input layer, several possible hidden layers, and the output layer. Each layer is connected to the next layer for transferring information \cite{schmidhuber2015deep}. In every single layer, there is no connection among its neurons. Therefore, we represent the information bypass from the $l-1$ layer to $l$ via the vector $\bm{z}^l$. Every component of vector $\bm{z}^l$ is computed by 
\begin{equation}
\label{eq:NN_1}
    {z}^l_m = {a} (\sum_{n=1}^{N_l} w^l_{mn} {z}_n^{l-1} + b^l_{m} ),~~~l=1,\ldots,L. 
\end{equation}
In Eq.\,(\ref{eq:NN_1}), ${z}^{l-1}_n$, is the $n$-th neuron within the $l-1$-th layer. The component $w_{mn}$ shows the connection weight between the $n$-th neuron of the layer $l-1$ and the $m$-th neuron of the layer $l$. Every individual neuron in the $l$-th hidden layer owns a bias variable $b_m^l$. The number $N_I$ corresponds to the number of neurons in the $l$-th hidden layer. The total number of hidden layers is $L$. The letter $a$ stands for the activation function in each neuron. The activation function $a(\cdot)$ is usually a non-linear function. In this work, the hyperbolic-tangent function is utilized which is defined as
\begin{equation}
\label{eq:af}
    \tanh(x)=\dfrac{\exp^x - \exp^{-x}}{\exp^x + \exp^{-x}}. 
\end{equation}
The proper choice of the activation function is problem dependent and shall be obtained based on hyperparameter studies on the final performance of the network. See also discussions in \cite{REZAEI2022PINN}, \cite{Jagtaop22} and \cite{Abbasi2022}.

The input layer of the PINNs for the described thermo-mechanical problem consists of the position vector $x$ and $y$ coordinates. For the output layer, we have $u_x$, $u_y$, $\sigma_{xx}$, $\sigma_{xy}$, and $\sigma_{yy}$ in a mechanical field and $T$, $q_x$, and $q_y$ in the thermal field. In the current work, the focus is on a 2-D setting, and extension to a 3-D setting is trivial. The rest of the network architecture will be discussed in what follows.
The trainable set of parameters of the network is represented by $\bm{\theta}=\{ \bm{W},\bm{b}\}$, which are the weights and biases of a neural network, respectively. Considering each neural network structure as $\mathcal{N}$, the final outcome of the network for the mechanical and thermal field reads
\begin{align}
\label{eq:NN_mech}
    u_x           = \mathcal{N}_{u_x} (\bm{X}; \bm{\theta}),~
    u_y           = \mathcal{N}_{u_y} (\bm{X}; \bm{\theta}),~
    \sigma^o_{xx} = \mathcal{N}_{\sigma_{xx}} (\bm{X}; \bm{\theta}),~
    \sigma^o_{xy} = \mathcal{N}_{\sigma_{xy}} (\bm{X}; \bm{\theta}),~
    \sigma^o_{yy} = \mathcal{N}_{\sigma_{yy}} (\bm{X}; \bm{\theta}),
\end{align}

\begin{align}
\label{eq:NN_ther}
    T      = \mathcal{N}_{T} (\bm{X}; \bm{\theta}),~
    q_x^o  = \mathcal{N}_{q_x} (\bm{X}; \bm{\theta}),~
    q_y^o  = \mathcal{N}_{q_x} (\bm{X}; \bm{\theta}).
\end{align}
The neural network outputs are functions of the trainable parameters and the training is done by minimizing different physical loss functions. Next, we build the BVP via the defined input and output layers. To do so, the partial differential equation and its corresponding initial and/or boundary conditions will be defined in terms of loss functions for the neural networks. Therefore, we require to obtain the derivatives of the output layer with respect to the input layer (for instance, the derivative of stress with respect to $x$- or $y$-direction). Current deep learning packages, i.e, Pytorch \cite{paszke2017automatic} and Tensorflow \cite{abadi2016tensorflow} are capable of computing the derivatives based on the automatic differentiation algorithms \cite{baydin2018automatic}. The algorithms developed in the current work are implemented using the SciANN package \cite{SciANN} but the methodology can easily be transferred to other platforms.

Denoting the summation of mechanical-related loss terms by $\mathcal{L}_M$, it is defined based on a set of equations, Eqs.\,(\ref{eps_t})\,-\,(\ref{BcsMech_n}), as
\begin{align}
\label{Totalloss_mech}
\mathcal{L}_M &= \underbrace{\mathcal{L}^{EF}_M + \mathcal{L}^{DBC}_M}_{\text{based on $\bm{u}$}} + \mathcal{L}^{cnc}_M + \underbrace{\mathcal{L}^{SF}_M  + \mathcal{L}^{NBC}_M}_{\text{based on $\bm{\sigma}^o$}}.
\end{align}
In Eq.\,(\ref{Totalloss_mech}), the energy form of the problem ($\mathcal{L}_M^{EF}$) and the prescribed Dirichlet boundary conditions losses ($\mathcal{L}_M^{DBC}$) are minimized by means of the displacement vector $\bm{u}$ output. For Neumann boundary conditions ($\mathcal{L}_M^{NBC}$) and the strong form ($\mathcal{L}_M^{SF}$) of the problem requiring the first-order displacement derivatives, relevant loss terms are applied to the $\bm{\sigma^o}$,
\begin{align}
\label{loss_weak}
\mathcal{L}^{EF}_M &= \text{MAE}^{EF}_M\left(\int_{\Omega}\dfrac{1}{2}\hat{\bm{\varepsilon}}_e^T\,\hat{{{C}}}(x,y)\,\hat{\bm{\varepsilon}}_e\,~dV-
\int_{\Gamma_N} {\bm{u}^T}\,\bar{\bm{t}}~dA\right), \\
\label{loss_DBC}
\mathcal{L}^{DBC}_M &= \text{MSE}^{DBC}_M\left( \bm{u} - \overline{\bm{u}} \right), \\
\label{loss_cnc}
\mathcal{L}^{cnc}_M &= \text{MSE}^{cnc}_M\left( \bm{\sigma}^o - \bm{\sigma} \right), \\
\label{loss_SF}
\mathcal{L}^{SF}_M &= \text{MSE}^{SF}_M\left( \text{div}(\bm{\sigma}^o) \right), \\
\label{loss_NBCx}  
\mathcal{L}^{NBC}_M &= \text{MSE}^{NBC}_M\left(\bm{\sigma}^o \cdot \bm{n} - \overline{\bm{t}} \right).
\end{align}
The loss function related to the thermal field is also computed based on Eqs.\,(\ref{StrongfromThermal})-\,(\ref{int_Energy_ther}). 

Analogously to the mechanical losses, the thermal loss $\mathcal{L}_T$ consists of $\mathcal{L}_T^{EF}$ (energy form of the thermal diffusion problem), $\mathcal{L}_T^{DBC}$ (Dirichlet boundary conditions), $\mathcal{L}_T^{NBC}$ (Neumann boundary conditions), $\mathcal{L}_T^{SF}$ (strong form of diffusion problem), and $\mathcal{L}_T^{cnc}$ (connection term) and it reads 
\begin{align}
\label{Totalloss_th}
\mathcal{L}_T &= \underbrace{\mathcal{L}^{EF}_T + \mathcal{L}^{DBC}_T}_{\text{based on $T$}} + \mathcal{L}^{cnc}_T + \underbrace{\mathcal{L}^{SF}_T  + \mathcal{L}^{NBC}_T}_{\text{based on $\bm{q}^O$}}, \\
\label{loss_weak_th}
\mathcal{L}^{EF}_T &= \text{MAE}^{EF}_T\left( \int_{\Omega}^{} \frac {1}{2} \bm{q}^T \cdot \nabla T \,dV + \int_{\bar{\Gamma}}^{} (\bm{q} \cdot \bm{n})~T \,dA \right), \\
\label{loss_DBC_th}
\mathcal{L}^{DBC}_T &= \text{MSE}^{DBC}_T\left( T - \overline{T} \right), \\
\label{loss_cnc_th}
\mathcal{L}^{cnc}_T &= \text{MSE}^{cnc}_T\left( \bm{q}^o - \bm{q} \right), \\
\label{loss_SF_th}
\mathcal{L}^{SF}_T &= \text{MSE}^{SF}_T\left( \text{div}(\bm{q}^o) \right), \\
\label{loss_NBC_th}  
\mathcal{L}^{NBC}_T &= \text{MSE}^{NBC}_T\left( \bm{q}^o \cdot \bm{n} - \overline{\bm{q}} \right).
\end{align}

For completeness, the mean squared and mean absolute error are defined as
\begin{align}
\label{eq:MSE}
\text{MSE}( \bullet )_{type} &= \dfrac{1}{k_{type}}\sum_{i=1}^{k_{type}}(\bullet)^2,\quad
\text{MAE}( \bullet )_{type} = \dfrac{1}{k_{type}}\sum_{i=1}^{k_{type}}|\bullet|.
\end{align}
The mathematical optimization problem is written as  
\begin{align}
\label{minimize}
\bm{\theta}^* = \arg \min_{\bm{\theta}} \mathcal{L}(\bm{X}; \bm{\theta}),
\end{align}
where $\bm{\theta}^*$ are the optimal trainable parameters (weights and biases) of the network which is the result of the optimization problem.\\[4pt] 
\textbf{Remark 1} The energy form Eqs.\,(\ref{loss_weak}) and (\ref{loss_weak_th}) are evaluated at the global level and a single loss term is available for all of the collocation points. The other loss terms are minimized at every single collocation point. As discussed in \cite{REZAEI2022PINN}, choosing MAE for the energy loss is beneficial to keep its value in the order of other loss terms. \\[4pt] 
\textbf{Remark 2} It should be noted that the stress outputs of the network are used in the strong form and Neumann boundary conditions losses. Therefore, potential problems related to being constrained by saddle points that can arise in mixed formulations such as the Hellinger-Reissner principle do not arise.
\\[4pt] 
We investigate two distinct training algorithms to solve the coupled system of equations. In one approach, we utilize a so-called sequential training, and in the other approach, we intend to use a coupled training as depicted in Fig.\,\ref{fig:intro}. The sequential training algorithm is done by minimizing losses of one individual physics while the other physical parameters are kept frozen. Readers are also encouraged to see \citet{HAGHIGHATAmini22} for more details. Sequential training has a similar procedure as staggered numerical schemes \cite{Ruan2022, REZAEI2022108177, Felippa1980Stag}. For sequential training, we define the following loss terms 
\begin{align}
\label{eq:Seq_loss_phys1}
\mathcal{L}_{Seq1}&=\mathcal{L}_{\text{Phy1}}=\mathcal{L}_T,\\
\label{eq:Seq_loss_phys2}    
\mathcal{L}_{Seq2}&=\mathcal{L}_{\text{Phy2}}=\mathcal{L}_M. 
\end{align}
During the minimization of the loss function in Eq.\,(\ref{eq:Seq_loss_phys1}), the output related to the mechanical field (i.e. ${u}_x, {u}_y, {\sigma}_{xx}^o, {\sigma}_{xy}^o,~\text{and}~{\sigma}_{yy}^o$) is kept frozen. While the loss function in Eq.\,(\ref{eq:Seq_loss_phys2}) is minimized, the output related to the thermal field (i.e. $T, {q}_x^o,~\text{and}~{q}_y^o$) is kept frozen. 
On the other hand, we have coupled training where we minimize all the loss terms, simultaneously. 

For the coupled training, the total loss function is the summation of all the loss terms and it is written as
\begin{align}
\label{eq:Coup_loss}
\mathcal{L}_{Cpl}&=\mathcal{L}_{\text{Phy1}}\,+\mathcal{L}_{\text{Phy2}}=\mathcal{L}_M+\mathcal{L}_T.
\end{align} 
%%%%%%%%%%%%%%%%%%

Fig.\,\ref{fig:PINNs_net_ML} depicts the mixed PINNs formulation for solving the thermoelasticity problem. In addition to the position vector ($x$, $y$), the material parameters $C(x,y)$ and $k(x,y)$ are added to the input layer. The latter avails the capability of a network to learn the solution for other material properties. 

In the proposed network structure, the primary variables and their spatial derivatives are directly evaluated. For example, when it comes to the mechanical problem, the displacement vector $\bm{u}$ and the stress tensor $\bm{\sigma^o}$ are defined as network output. Moreover, the strain tensor $\bm{\varepsilon}$ is calculated by means of the predicted values for deformation ($u_x$ and $u_y$), and the predicted temperature $T$ according to Eqs.\,(\ref{kinematics}) and (\ref{eps_t}). By inserting the computed strain tensor ($\bm{\varepsilon}$) into Eq.\,(\ref{materiallaw}), one computes the stress tensor $\bm{\sigma}$ for the coupled system of equations. Consequently, the resulting stress tensor from the evaluated displacement $\bm{\sigma}$ is linked to the network's direct evaluation $\bm{\sigma}^o$ through the connection loss term $\mathcal{L}^{cnc}_M$ in Eq.\,(\ref{loss_cnc}). The latter is a soft constraint to connect $\bm{\sigma}^o$ to $\bm{\sigma}$ values. 

The analogous strategy is employed for the thermal field in which the heat flux vector $\bm{q}$ is derived on the basis of Eq.\,(\ref{Fourier}) where we compute the thermal field gradients by means of the automatic differentiation. The $\bm{q}$ is connected to $\bm{q}^o$ by the $\mathcal{L}
_T^{cnc}$ described in Eq.\,(\ref{loss_cnc_th}).

According to Fig.\,\ref{fig:PINNs_net_ML}, the optimization problem in a sequential training procedure starts with the first desired field loss function to enforce the minimization of the first physical field loss functions.
\vspace{-3mm}
In the thermoelasticity problem, the training starts by finding the optimum network parameters to satisfy the thermal loss function ($\mathcal{L}_T$). This choice is made based on the fact that the thermal-induced strain tensor will affect the results of the mechanical field. In our proposed algorithm, the training is stopped for the thermal field loss function (first physics' loss function) after reaching a certain number of epochs $n_T$. Subsequently, the training is done for the next physics' loss function (mechanical field loss function $\mathcal{L}_m$) for the same or a different number of epochs $n_M$. The whole training will be stopped if the total number of epochs is reached the desired value $n_A$. While the desired value for the total number of epochs has not been reached, the same procedure will be continued to reach that particular value (minimizing $\mathcal{L}_T$ for $n_T$ epochs and minimizing $\mathcal{L}_M$ for $n_M$ epochs).
For the coupled training procedure, the total loss function is computed by summing the different physics loss functions ($\mathcal{L}_M$ and $\mathcal{L}_T$) into a single one $\mathcal{L}_{Cpl}$, see Eq.\,(\ref{eq:Coup_loss}) and Fig.\,\ref{fig:PINNs_net_ML}. 
\\ \\
\textbf{Remark 3} It is shown that employing separated networks and utilizing only first-order derivatives in the network's loss function leads to a better performance \cite{REZAEI2022PINN}. See also studies by \citet{CHIU2022} on this matter. Reducing the order of derivatives also has other benefits. For example, one has more flexibility in choosing the activation functions (i.e. we avoid problems related to vanishing gradients due to higher-order derivatives). Moreover, the computational cost is decreased compared to the cases where we have second-order derivation.
\\ \\
\textbf{Remark 4} Another possibility to stop each training procedure in the sequential training is to reach a certain desired value of the loss for each field. This is achieved by having an infinite number for the number of epochs while optimizing each field's loss function. The training procedure will stop whenever a certain value for the desired loss function is accomplished. 
\\ \\
\textbf{Remark 5} The current mixed PINNs approach can also be interpreted as a multi-objective optimization problem. Therefore, having a proper balance between different loss terms is essential. Often in multi-physical problems, it is important to normalize the quantities before the minimization starts \cite{Henkes2022}. Also, the loss terms $L_T^{EF}$ and $L_M^{EF}$ are defined based on the means of absolute error to be in the same order as other loss terms.  
\\ \\
In this work, the Adam optimizer is employed, and the relevant network parameters are summarized in Table\,\ref{tab:network}. Please note that in the case of a simple PINNs formulation, where the energy form of the problem is absent, one can use multiple batches for the training of the network. 

\begin{table}[H]
\centering
\caption{Summary of the network parameters for coupled and sequential training procedures.}  
\label{tab:network}
\begin{footnotesize}
\begin{tabular}{ l l }
\hline
Parameter                          &  Value    \\
\hline
Input, Output                  &  $\bm{X}$, ($\bm{u}$, $\bm{\sigma}^o$, ${T}$, $\bm{q}^o$) \\ 
activation function                &  $\tanh$ \\ 
number of layers and neurons per layer ($L$, $N_l$)  &  (5, 40) \\
batch size                         &  full batch for the mixed PINNs formulation  \\ 
(learning rate $\alpha$, number of epochs)  &  $(10^{-3},10^{5})$ \\ 
\hline 
coupled training \\
total number of epochs $(n_A)$ & $100$k epochs \\
\hline
sequential training \\
total number of iterations $(n_A)$ & $200$k epochs \\
iterations for the thermal field before switch $(n_T)$ & $20$k epochs \\
iterations for the mechanical field before switch $(n_M)$ & $20$k epochs\\
\hline
\end{tabular}
\end{footnotesize}
\end{table}

%%%%%%%%%%%%%%%%%%%%%%%%
Based on the boundary conditions discussed in Fig.~\ref{fig:BC}, we write all the loss terms more specifically for the given BVP in what follows. 
The components of the total strain tensor and temperature gradient vector are calculated by taking derivatives of displacements and temperature fields, respectively. Considering Eqs.\,(\ref{eq:NN_mech}) and (\ref{eq:NN_ther}), we define the following networks  
\begin{align}
\mathcal{N}_{\varepsilon_x}(\bm{X};\bm{\theta}) &= \partial_x \mathcal{N}_{u_{x}},\\
\mathcal{N}_{\varepsilon_y}(\bm{X};\bm{\theta}) &= \partial_y \mathcal{N}_{u_{y}},\\
\mathcal{N}_{\varepsilon_{xy}}(\bm{X};\bm{\theta}) &= \dfrac{1}{2}\left(\partial_{y} \mathcal{N}_{u_{x}} + \partial_{x} \mathcal{N}_{u_{y}} \right),\\
\mathcal{N}_{\nabla_xT}(\bm{X};\bm{\theta}) &= \partial_x\mathcal{N}_T,\\
\mathcal{N}_{\nabla_yT}(\bm{X};\bm{\theta}) &= \partial_y\mathcal{N}_T.
\end{align}
Components of the (isotropic) thermal strain tensor are then derived by 
\begin{align}
\mathcal{N}_{\varepsilon_{t_x}}(\bm{X};\bm{\theta})=\mathcal{N}_{\varepsilon_{t_y}} &= \alpha\left(\bm{X}\right) \left(\mathcal{N}_{T}-T_0\right).
\end{align} 
Considering the total and thermally induced strain tensors, the elastic part of the strain tensor reads 
\begin{align}
\mathcal{N}_{\varepsilon_{e_x}}(\bm{X};\bm{\theta}) &= \mathcal{N}_{\varepsilon_x}-\mathcal{N}_{\varepsilon_{t_x}},\\
\mathcal{N}_{\varepsilon_{e_y}}(\bm{X};\bm{\theta}) &=\mathcal{N}_{\varepsilon_y}-\mathcal{N}_{\varepsilon_{t_y}},\\
\mathcal{N}_{\varepsilon_{e_{xy}}}(\bm{X};\bm{\theta}) &= \mathcal{N}_{\varepsilon_{{xy}}}.
\end{align}
Utilizing the constitutive relations (Hooke's law) as well as Fourier's law, one computes the components of the stress tensor and heat flux vector according to the following relations
\begin{align} %\begin{split}
 \mathcal{N}_{\sigma_x}(\bm{X};\bm{\theta}) &= \dfrac{E(\bm{X})}{(1+\nu(\bm{X}))(1-2\nu(\bm{X}))}\left((1-\nu(\bm{X}))\mathcal{N}_{\varepsilon_{e_x}}+\nu(\bm{X})  \mathcal{N}_{\varepsilon_{e_y}} \right) , \\
 \mathcal{N}_{\sigma_y}(\bm{X};\bm{\theta}) &= \dfrac{E(\bm{X})}{(1+\nu(\bm{X}))(1-2\nu(\bm{X}))}\left(\nu(\bm{X})\mathcal{N}_{\varepsilon_x}+(1-\nu(\bm{X}))  \mathcal{N}_{\varepsilon_y} \right), \\
 \mathcal{N}_{\sigma_{xy}}(\bm{X};\bm{\theta}) &= \dfrac{E(\bm{X})}{2(1+\nu(\bm{X}))}\mathcal{N}_{\varepsilon_{e_{xy}}},\\
 \mathcal{N}_{q_x}(\bm{X};\bm{\theta}) &= -k(\bm{X})\mathcal{N}_{\nabla_xT},\\
\mathcal{N}_{q_y}(\bm{X};\bm{\theta}) &= -k(\bm{X})\mathcal{N}_{\nabla_yT}. %\end{split}
\end{align}

To simplify equations for formulating loss functions, in what follows, the dependency of neural network output $\mathcal{N}\left(\bm{X};\bm{\theta}\right)$ on the collocation points' coordinates $\bm{X}$ and trainable parameters $\bm{\theta}$ are not shown. 

Finally, the mathematical expressions for calculating loss functions for the mechanical and thermal field are expanded in what follows. See also Eqs.\,(\ref{loss_weak}) to (\ref{loss_NBC_th}) as well as Fig.~\ref{fig:PINNs_net_ML},

\begin{align}
\label{loss_weak_exp} %%%%%%%%%%%%%%%
\mathcal{L}^{EF}_M &= \left| - \dfrac{1}{2 N_{b}}\sum_{\bm{X}\in \Omega} 
\left( 
\mathcal{N}_{\sigma_x} \mathcal{N}_{\varepsilon_{e_x}} + \mathcal{N}_{\sigma_y} \mathcal{N}_{\varepsilon_{e_y}} +
\mathcal{N}_{\sigma_{xy}} \mathcal{N}_{\varepsilon_{e_{xy}}} \right)\right|, \\ \nonumber
\label{loss_DBC_exp} %%%%%%%%%%%%%%%%%
\mathcal{L}^{DBC}_M  &= \dfrac{1}{N^L_{DBC}} \sum_{\bm{X}\in \Gamma^L_D} \left(\mathcal{N}_{u_{x}}-0\right)^2 + 
\dfrac{1}{N^R_{DBC}} \sum_{\bm{X}\in \Gamma^R_D}  \left(\mathcal{N}_{u_{x}}-0\right)^2, + 
\dfrac{1}{N^T_{DBC}} \sum_{\bm{X}\in \Gamma^T_D}
\left(\mathcal{N}_{u_{y}}-0\right)^2 \\ 
&+ 
\dfrac{1}{N^B_{DBC}} \sum_{\bm{X}\in \Gamma^B_D} \left(\mathcal{N}_{u_{y}}-0\right)^2, \\ 
\label{loss_cnc_exp} %%%%%%%%%%%%%%%%
\mathcal{L}^{cnc}_M&= \dfrac{1}{N_{b}} \sum_{\bm{X}\in \Omega} \left( \left(\mathcal{N}_{\sigma_{x}^o} - \mathcal{N}_{\sigma_{x}}\right)^2  + 
\left(\mathcal{N}_{\sigma_{y}^o} - \mathcal{N}_{\sigma_{y}}\right)^2  +
 \left(\mathcal{N}_{\sigma_{xy}^o} - \mathcal{N}_{\sigma_{xy}}\right)^2
\right), \\
\label{loss_SF_exp} %%%%%%%%%%%%%%%%%%
\mathcal{L}^{SF}_M &= \dfrac{1}{N_{b}} \sum_{\bm{X}\in \Omega} \left( \left(\partial_x \mathcal{N}_{\sigma_{x}^o}\,+\, \partial_y \mathcal{N}_{\sigma_{xy}^o}\right)^2  + 
\left(\partial_y \mathcal{N}_{\sigma_{y}^o}\,+\, \partial_x \mathcal{N}_{\sigma^o_{xy}}\right)^2 
\right), \\ 
\label{loss_NBCx_exp} %%%%%%%%%%%%%%%%
\mathcal{L}^{NBC}_M &= 0.
\end{align}

In the above equations, $\Gamma^L$, $\Gamma^R$, $\Gamma^T$, and $\Gamma^B$ denote the location of points at left, right, top, and bottom boundaries, respectively. \emph{DBC} and \emph{NBC} represent for Dirichlet and Neumann boundary conditions. 

Note, that the loss term related to the Neumann BCs (Eq.\,(\ref{loss_NBCx_exp})) is zero according to the defined BVP. We do not apply any external traction to the microstructure of the material.
For the computation of the integral in the energy form, we use the summation over the collocation points with equal weights as the collocation points are uniformly distributed in the domain. Therefore, $N_{b}$ denotes the number of collocation points within the domain (body). Moreover, $N^L$, $N^R$, $N^T$, and $N^B$ represent the number of collocation points at left, right, top, and bottom boundaries, respectively. 
See also \cite{FUHG2022110839} for other possible methodologies. 

The loss functions related to the thermal field read as
\begin{align}
\label{loss_EF_T}
\mathcal{L}^{EF}_T &= 
\left| 
\dfrac{1}{2 N_b}\sum_{\bm{X}\in \Omega}
\left( 
\mathcal{N}_{q_x} \mathcal{N}_{\nabla_x T} + \mathcal{N}_{q_y} \mathcal{N}_{\nabla_y T} \right) + 
\dfrac{1}{N_{NBC}^L} \sum_{\bm{X}\in \Gamma_N^L}
\left(\mathcal{N}_{q_x} \mathcal{N}_T \right)
\right|, \\
\mathcal{L}^{DBC}_T &= \dfrac{1}{N^L_{DBC}} \sum_{\bm{X}\in \Gamma^L_D} \left(\mathcal{N}_{T}-1\right)^2 + 
\dfrac{1}{N^R_{DBC}} \sum_{\bm{X}\in \Gamma^R_D} \left(\mathcal{N}_{T}-0\right)^2, \\
\mathcal{L}^{cnc}_T &= \dfrac{1}{N_{b}} \sum_{\bm{X}\in \Omega} \left( \left(\mathcal{N}_{q_{x}^o} - \mathcal{N}_{q_{x}}\right)^2  + 
\left(\mathcal{N}_{q_{y}^o} - \mathcal{N}_{q_{y}}\right)^2 \right), \\
\mathcal{L}^{SF}_T &= \dfrac{1}{N_{b}} \sum_{\bm{X}\in \Omega} \left(\partial_x \mathcal{N}_{q_{x}^o}\,+\, \partial_y \mathcal{N}_{q^o_{y}}\right)^2 ,\\ \nonumber
\mathcal{L}^{NBC}_T &= \sum_{\bm{X}\in \Gamma^T_N} \dfrac{1}{N^T_{NBC}}\left(\mathcal{N}_{q^o_{y}}-0\right)^2 \\
&+ \sum_{\bm{X}\in \Gamma^B_N}\dfrac{1}{N^B_{NBC}}\left(\mathcal{N}_{q^o_{y}}-0\right)^2.
\end{align}

%%%%%%%%%%%%%%%%%%%%
%%%%%%%%%%%%%%%%%%%%
\newpage
\section{Results}
\subsection{A circular heterogeneity in a matrix (geometry 1)} \label{sec:1st_example}
\begin{figure}[H] 
  \centering
  \includegraphics[width=1.0\linewidth]{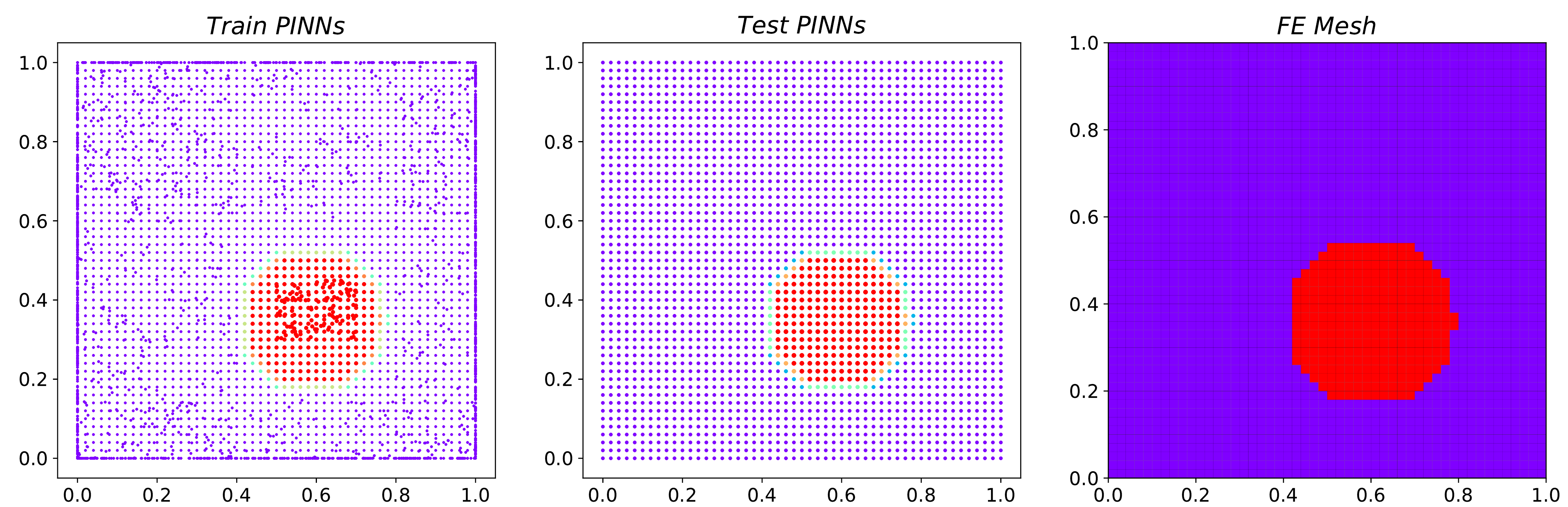}
  \caption{Left: collocation points for minimizing the loss functions in PINNs. Middle: points for evaluating the trained network response. Right: discretized domain using FEM to solve the same boundary value problem in thermoelasticity. The color variation shows the different values of Young's modulus.}
  \label{fig:geom1_config}  
  
\end{figure}

We start with the first geometry which is described in Fig.\,\ref{fig:allgeom}. The collocation points are set as depicted on the left-hand side of Fig.\,\ref{fig:geom1_config}. As explained in \cite{REZAEI2022PINN}, to enhance the accuracy of the results, one needs to increase the collocation points in the heterogeneity, domain boundaries, and matrix. Approximately, $5000$ collocation points are considered in a rectangular domain from $0$ to $1$\,[mm] in both $x$- and $y$-directions. Minimization of the loss functions (described in Sec.\,\ref{sec:thermos}) is done at the collocations points. The material parameters for the current boundary value problem are mentioned in Tab.\,\ref{tab:par1}.

\begin{table}[H]
	\centering
	\begin{tabular}{ l l  } \hline
	\multirow{1}{*}{}         & Value/Unit    \\ \hline \hline 
	\textbf{Thermal elasticity problem of geometry 1} \\
	 Matrix Young's modulus and Poisson's ratio ($E_{\text{mat}}$, $\nu_{\text{mat}}$)  & ($0.3$ GPa, $0.3$)  \\
 	Inclusion Young's modulus and Poisson's ratio ($E_{\text{inc}}$, $\nu_{\text{inc}}$)  & ($1.0$ GPa, $0.3$)  \\
 	Matrix heat conductivity (${k}_{\text{mat}}$)  & $0.3$~W/mK  \\
   	Inclusion heat conductivity (${k}_{\text{inc}}$)  & $1.0$~W/mK  \\
    \hline
	\end{tabular}
	\caption{Geometry 1 input parameters for the linear thermal elasticity problem}
	\label{tab:par1}
\end{table}
 
To evaluate the network response, the thermoelasticity problem is also solved for the same domain by the finite element program FEAP \cite{Feap}. The discretization of the domain for FE analysis is done in a way that the elements coincide with the regular collocation points for evaluating the response of the trained network. 
Moreover, in Fig.\,\ref{fig:geom1_config}, the variation of color shows the different Young's moduli and thermal conductivity values in the PINNs collocation points. The latter comes from the fact that some points are living at the interface of the matrix and the inclusion. The same procedure exists during the assembly procedure in the FE method.
The minimization of the loss functions is done by two different procedures, coupled and sequential training (see also Fig.\,\ref{fig:PINNs_net_ML}). The training is performed on the Skylake CPU, Platinum 8160 model, and Intel HNS2600BPB Hardware Node Type for both cases. For the coupled training, every loss term is minimized by performing $100\,\text{k}$ epochs (iterations). To have a fair comparison for the sequential training, we perform a total of $200 \,\text{k}$ epochs. In the sequential training, we minimize each loss term for $100\,\text{k}$ epochs. The number of epochs is chosen based on the final loss value and accuracy of the results. This number may easily change based on the optimizer type and the preference for the desired accuracy.

%-------------------------------------------------------------------------------------
%%% coupled training tanh 
\subsubsection{Coupled training for the geometry 1}
\begin{figure}[H] 
  \centering
  \includegraphics[width=0.9\linewidth]{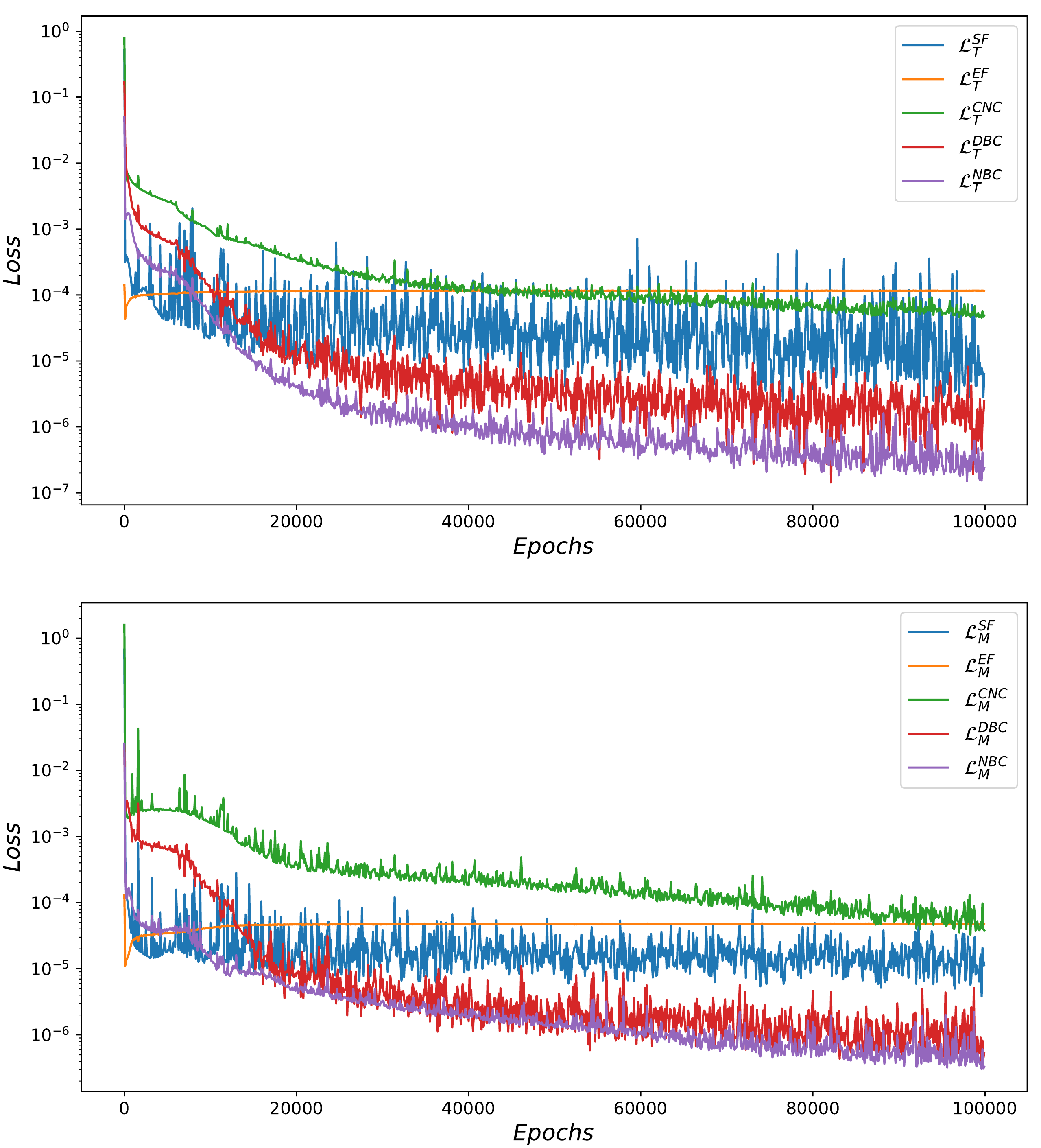}
  \caption{Prominent loss terms for the thermal and mechanical field decay during the training procedure utilizing the coupled training algorithm.   }
  \label{fig:losses_geo1_cup}
\end{figure}

The decay of prominent loss terms during the training procedure is shown in Fig.\,\ref{fig:losses_geo1_cup}.
In the coupled training minimization, we try to find the optimum network parameters to satisfy all of the loss terms, simultaneously.
Note that for the sake of clarity and a better comparison with the sequential training, the loss terms related to the mechanical and thermal sub-problem are plotted individually in Fig.\,\ref{fig:losses_geo1_cup}.
All the loss terms associated with governing equations and boundary conditions decrease as the number of epochs increases. It is interesting to note that the loss terms related to energy tend to remain more or less constant after a few thousand epochs. This observation may indicate that this term does not contribute significantly to the training process. However, we have observed that omitting this term results in poor outcomes compared to results from FEM. Instead, we believe that the energy of the system should be minimized, although it does not necessarily have to vanish at the equilibrium point. This is different from other loss terms, which are direct equations, that must be satisfied. Readers are also referred to \cite{REZAEI2022PINN} for a more detailed comparison between different approaches.

The deformed configurations of PINNs and FEM are plotted in
Fig.\,\ref{fig:defo_cop_tanh_5_40}. All the deformations are induced due to the introduced temperature gradient. To have a more detailed view, the comparison is made in Fig.\,\ref{fig:mech_cop_tanh_5_40} for the displacements and the stress components in PINNs and FEM results. The PINNs results are presented in the first column while the FEM results and the difference between PINNs and FEM are shown in the second and third columns, respectively. 

%\newpage
{Fig.\,\ref{fig:mech_cop_tanh_5_40} shows that the mechanical field results obtained from PINNs for the thermoelasticity problem are in good agreement with the outcomes from FE analysis. The average relative difference is $0.59\%$ for displacements in the $x$-direction and $0.87\%$ for stresses in the $x$-direction. While the maximum relative difference for the displacements and stress components is higher, these differences only occur at a few points in the domain near the inclusion boundaries. Local refinement of the collocation points can further improve the accuracy in these areas  \cite{REZAEI2022PINN}. 
\begin{figure}[H] 
  \centering
  \includegraphics[width=0.99\linewidth]{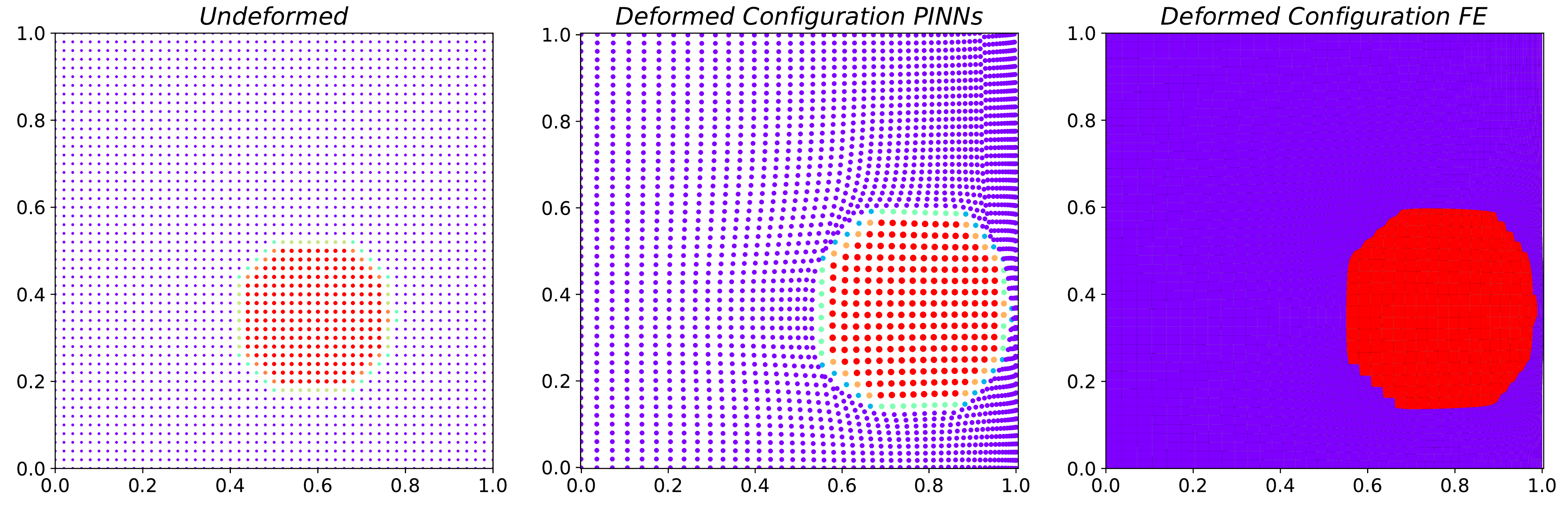}
  \caption{Left: undeformed configuration of evaluation points in PINNs. Middle: deformed configuration of evaluation points in PINNs employing coupled training. Right: deformed configuration of the discretized domain in FE analysis.}
  \label{fig:defo_cop_tanh_5_40}
\end{figure}

The maximum amount of difference is observed at the interface of the matrix and heterogeneity. This point is the result of having a rather sharp transition in the material properties at the interface. One possible remedy to overcome the latter problem is to add more collocation points near the interface or by applying domain decomposition methods, see \cite{Henkes2022, SHUKLA2021, KHARAZMI2021}. Increasing the number of epochs for training is also helpful.

The agreement of the thermal field results between PINNs and FE analysis in the thermoelasticity problem for the coupled training is also addressed in Fig.\,\ref{fig:tem_cop_tanh_5_40}. The relative averaged error for $T$ is about $0.3\,\%$ and for $q_x$ is around $0.76\,\%$. Similar to the mechanical field, the maximum relative difference occurs mostly at a few points on the interface. 

\begin{figure}[H] 
  \centering
  \includegraphics[width=0.95\linewidth]{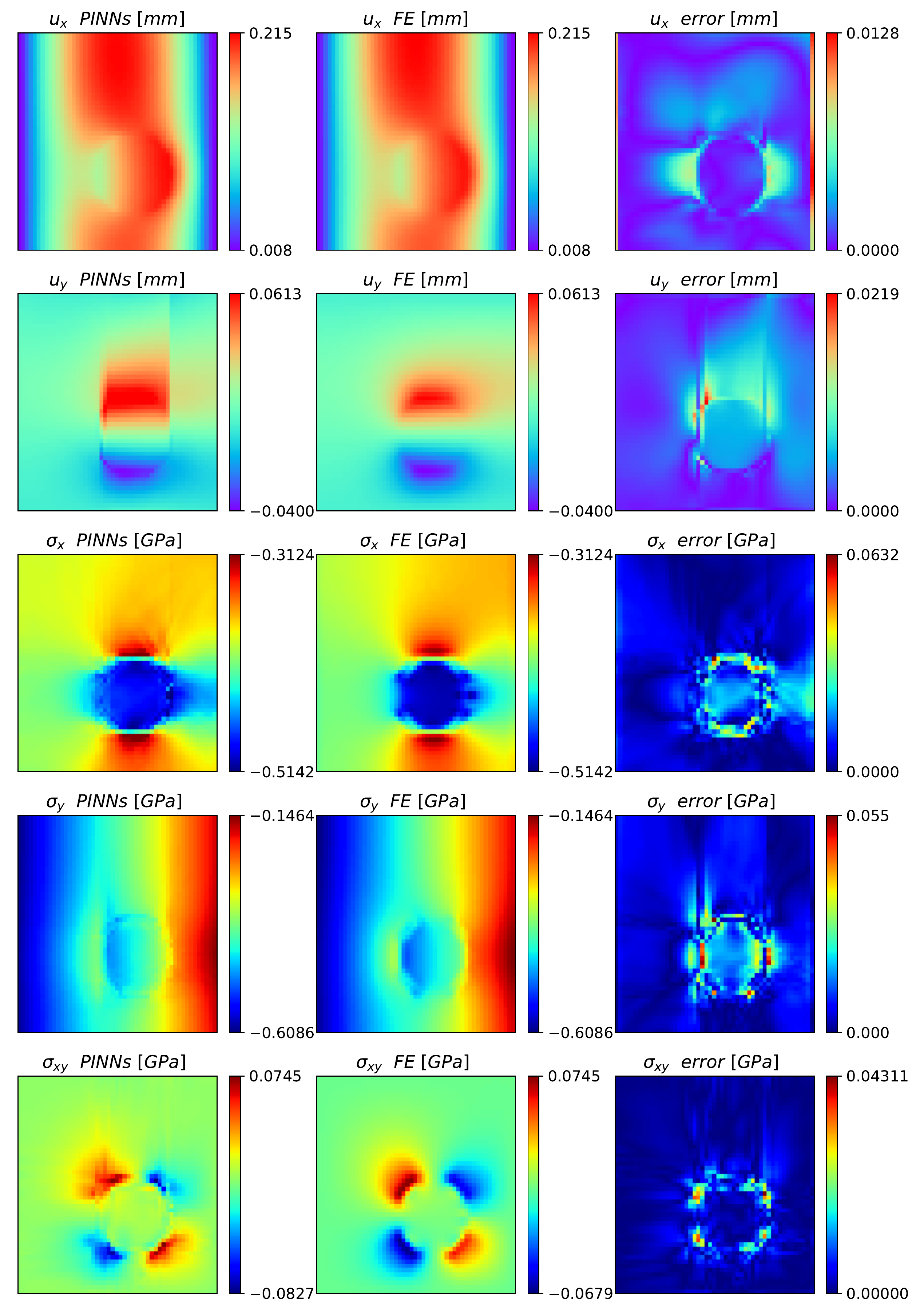}
  \caption{The mechanical field results of the thermoelasticity problem employing the coupled training procedure. Left: PINNs, Middle: FEM, Right: FEM with PINNs difference. }
  \label{fig:mech_cop_tanh_5_40}
\end{figure}

\begin{figure}[H] 
  \centering
  \includegraphics[width=0.88\linewidth]{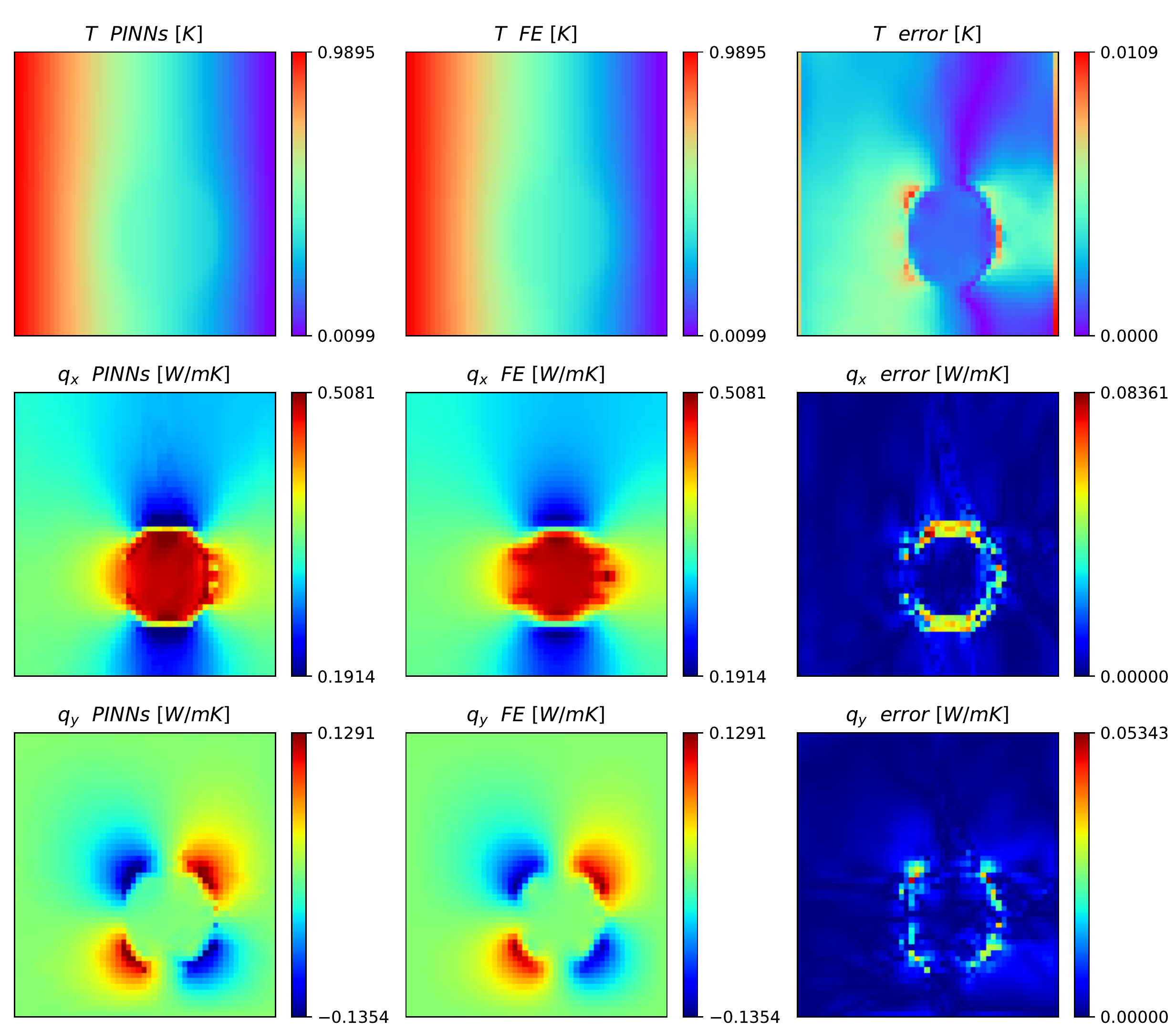}
  \caption{The thermal field results of the thermoelasticity problem employing the coupled training procedure. Left: PINNs, Middle: FEM, Right: FEM with PINNs difference. }
  \label{fig:tem_cop_tanh_5_40}
\end{figure}

%-------------------------------------------------------------------------------------

%%% Sequential training tanh 
\subsubsection{Sequential training}
The identical collocation points from Fig.\,\ref{fig:geom1_config} are utilized to minimize the thermal and mechanical loss functions ($\mathcal{L}_T$ and $\mathcal{L}_M$) by employing the sequential algorithm. As it is discussed beforehand, the minimization is performed in total for $200\,\text{k}$ epochs which results in $100\,\text{k}$ for each field loss function. The decay of main losses for each field with respect to the number is plotted separately for the mechanical and thermal fields in Fig.\,\ref{fig:losses_seq_tanh_5_40}.

In Fig.\,\ref{fig:losses_seq_tanh_5_40}, one can see the effect of chosen values for $n_T$ and $n_M$. These two numbers determine when we switch between the minimization of thermal and mechanical loss functions. The value for these two parameters is set to $20\,\text{k}$. Take into consideration that the loss terms for the thermal field are not evaluated during the mechanical part training, and vice versa. Therefore, to show the whole loss functions the losses are saved for each individual part during its training and later on appended together. One interprets that by minimization of the one field loss functions, other fields' losses may increase. The network tries to find the optimum parameters for the active field that may change the previously found optimum parameters for the nonactive fields. The latter point is hardly seen for each loss function right after upright lines in Fig.\,\ref{fig:losses_seq_tanh_5_40}, specifically for the thermal field losses. Finally, one needs to state that having some iterations between the fields during training is essential to find the global minimum of the system. For this example, $n_M$, $n_T$, and $n_A^{Seq}$ are chosen in a way to make five iterations possible. Further studies are required to find the optimum switching times between the fields and the number of epochs while minimizing one field loss function.
\begin{figure}[H] 
  \centering
  \includegraphics[width=0.9\linewidth]{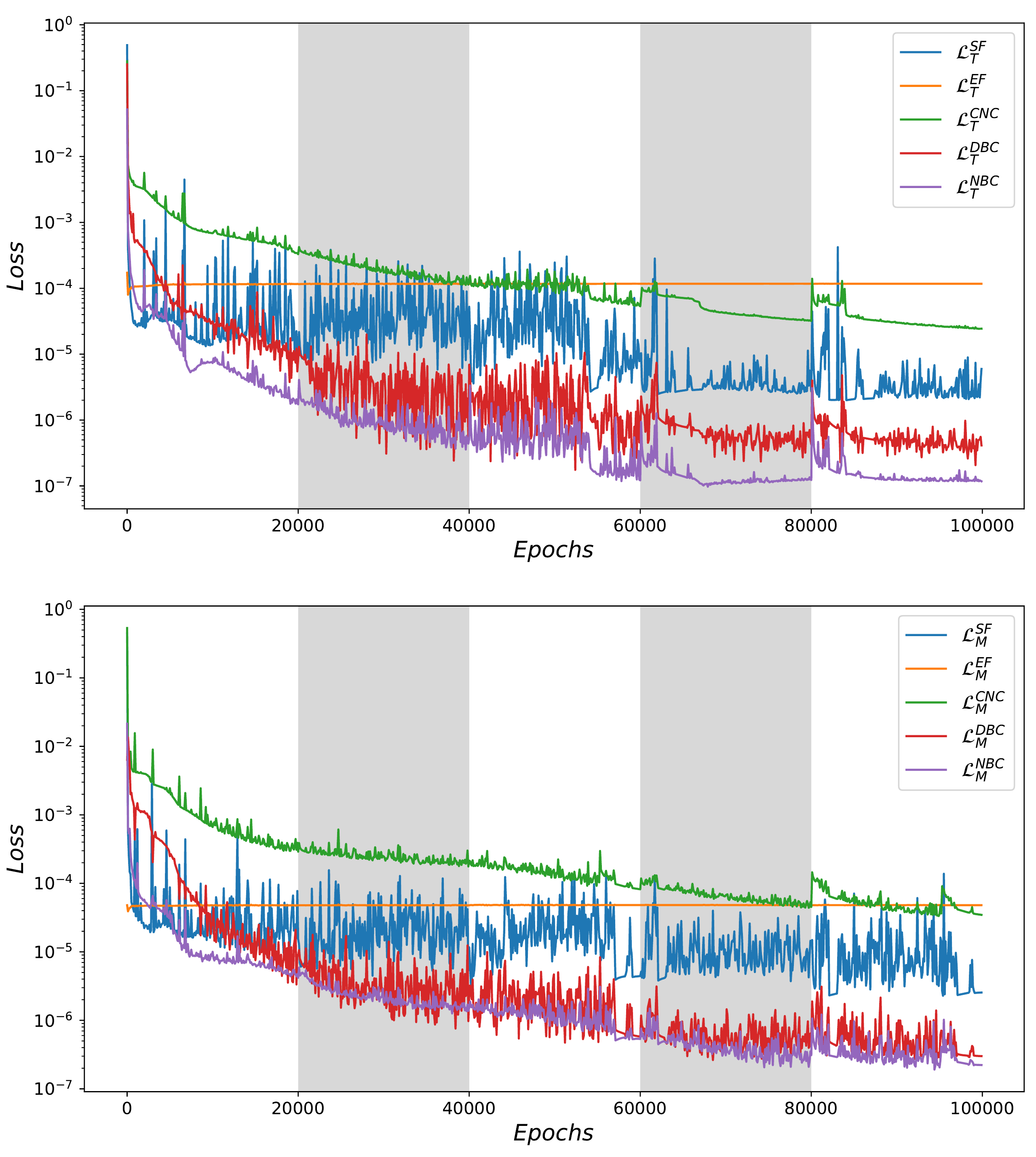}
  \caption{Top: prominent loss terms for the thermal field decay during the training procedure utilizing the sequential training algorithm. Bottom: prominent loss terms for the mechanical field decay during the training procedure utilizing the sequential training algorithm.  }
  \label{fig:losses_seq_tanh_5_40}
\end{figure}

Utilizing mixed PINNs by employing sequential training leads to $0.81\,\%$ and $0.8\,\%$ averaged relative difference for the displacement and stress components in $x$-direction by FE analysis, respectively. In the thermal field (see Fig.\,\ref{fig:tem_seq_tanh_5_40}), the averaged relative errors lie at $0.26\,\%$, and $0.66\,\%$ for the temperature field and the heat flux $q_y$ in $y$-direction.

Next, we compare the performance of our proposed mixed PINNs formulation to the standard PINNs structure \cite{RAISSI2019}, as well as the deep energy method (DEM) \cite{SAMANIEGO2020112790}, which uses only the energy term for training the neural network. We apply both the standard PINNs and DEM to solve the thermoelasticity boundary value problem with an equal number of training epochs and the same optimizer to have a fair comparison.
It is worth noting that the standard PINNs structure utilizes a separate neural network to predict the output variables $\bm{u}$ and $T$, which leads to second-order derivatives appearing in the formulation. The same architecture is used for the DEM method where thanks to the energy term only first-order derivatives are required. This network includes five hidden layers, each containing forty neurons with a \emph{tanh} activation function. We apply both the coupled and sequential training algorithms to train each of the standard PINNs, DEM, and mixed PINNs formulations.
 
\begin{figure}[H] 
  \centering
  \includegraphics[width=0.99\linewidth]{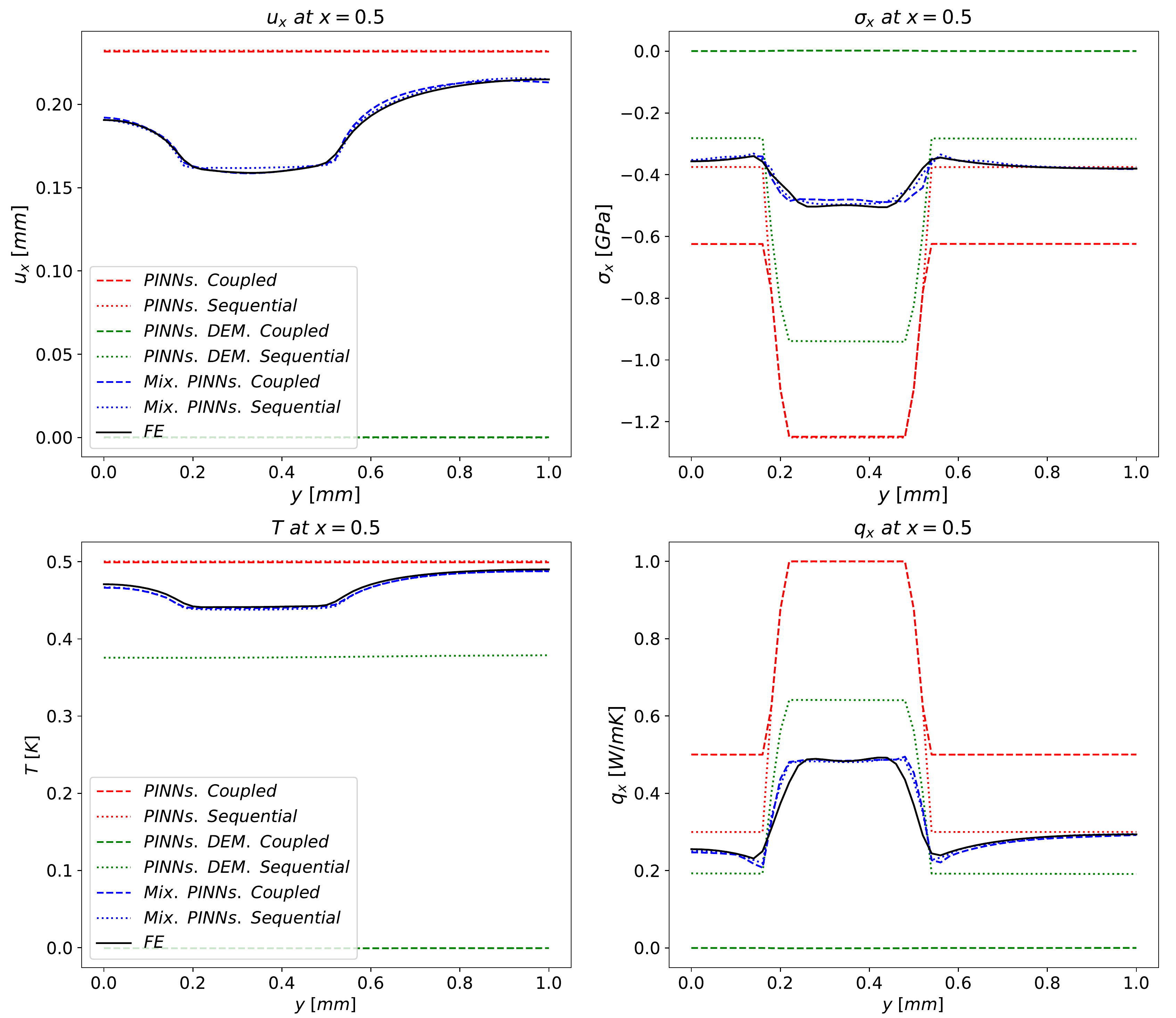}
  \caption{Comparison of the main output of the networks through the section at $x=0.5 \text{mm}$. Mix. PINNs: denotes the formulation of this paper by having the $\bm{\sigma}$ and $\bm{u}$ as output and utilizing the energy form loss in addition to conventional losses. PINNs: stands for the standard formulation that typically includes only the displacement field $\bm{u}$ and temperature field $T$ as outputs, with the strong form of the governing equations represented in the loss function. In contrast, the Deep Energy Method (DEM) employs the energy form of the problem in the loss function, also with $\bm{u}$ and $T$ as outputs, but without necessarily enforcing the strong form of the equations. Coupled: minimizing all fields loss functions simultaneously. Sequential: minimization is done with the proposed sequential training method.} 
  \label{fig:sectiongeo1}
\end{figure}

The $x$-components of displacement, stress, and heat flux as well as the temperature field are depicted in Fig.\,\ref{fig:sectiongeo1} for the section which is made at $x=0.5\,\text{mm}$. The FE analysis results are also added as a reference solution to the figures.

The standard PINNs formulation is unable to capture the effect of heterogeneity for the displacement component $u_x$ and the temperature field $T$. Additionally, it exhibits poor performance in predicting $q_x$ and $\sigma_x$, regardless of whether coupled or sequential training procedures are used. However, sequential training does yield slightly better results for stress and flux components. Similar observations can be made for DEM, where the results are better than those of standard PINNs but still fall short of our reference solution using FE, indicating the limitations of these methods for heterogeneous domains.
In contrast, the mixed PINNs formulation demonstrates much better performance and is capable of capturing the effect of heterogeneity using both coupled and sequential training algorithms. Upon closer examination of the results, we note that the coupled training algorithm has a slightly better performance than the sequential one. Moreover, regardless of the methodology used to minimize all the loss terms, we obtained better performance using sequential training over coupled training.
Overall, our findings indicate that the mixed PINNs formulation surpasses the standard PINNs and DEM methods in terms of accuracy and computational efficiency, making it a promising approach for solving complex physical problems. The latter is achieved by not only formulating the loss function solely based on first-order derivatives but also by using both the strong and weak forms of the problem and predicting the outputs via separate NN.  
\\ \\
\textbf{Remark 6} The average time of an epoch for coupled training is about $0.67$ seconds. For sequential training, this value decreased to $0.16$ seconds. However, since one needs to perform at least two iterations to achieve the solution in the case of sequential training, the computational gain is about $0.35$ seconds which reduces the computational time more than two times compared to the coupled training method. 
%%%%%%%%%%%%%%%%%%%%%%%%%%%%%%%%%%%%%%%%%%%%%%%
%\newpage
%%%%%%%%%%%%%%%%%%%%%%%%%%%%%%%%%%%%%%%% 
\subsection{Complex microstructure containing rectangular inclusions (geometry 2)}
\label{sec:comp_micro}
To further evaluate the performance of the proposed approach for solving multi-physical in the heterogeneous domains, the second geometry in Fig.\,\ref{fig:gem2} is now under consideration. The pattern of heterogeneity is defined to be more challenging with sharp corners. Such a pattern also corresponds to some specific material microstructures such as nickel alloy systems, see \cite{NGUEJIO2019}.
\begin{figure}[H] 
  \centering
  \includegraphics[width=0.99\linewidth]{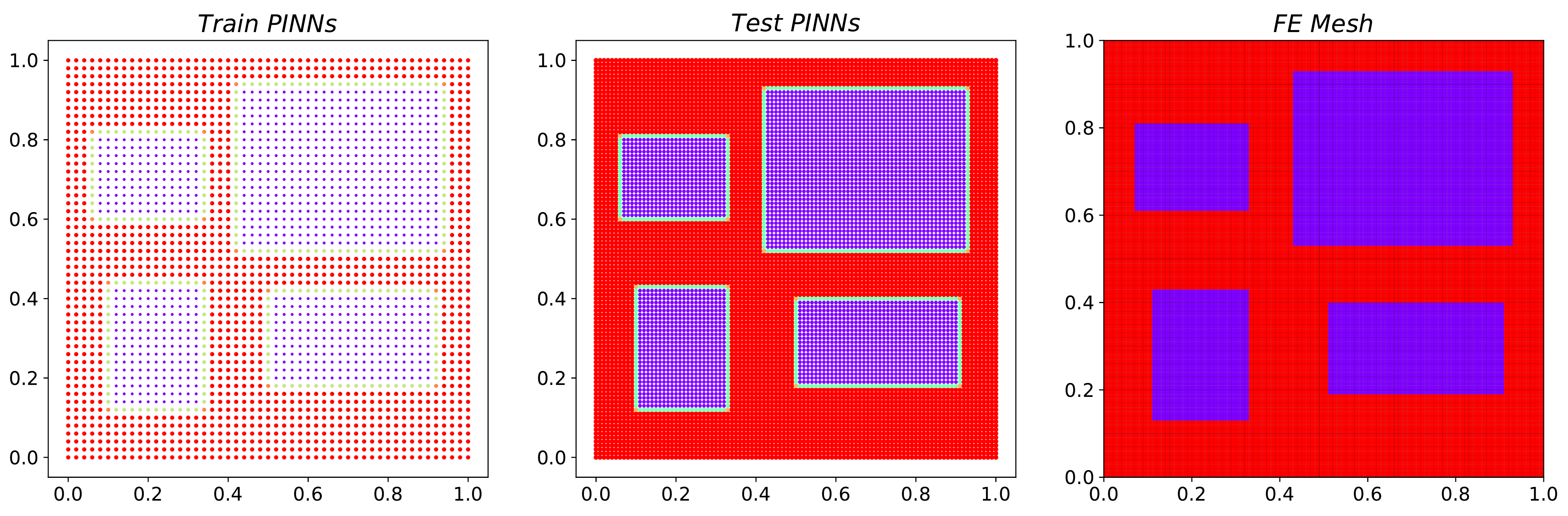}
  \caption{Left: collocation points for minimizing the loss functions for the second geometry. Middle: points for evaluating the trained network response. Right: discretized domain using FEM to solve the same boundary value problem in thermoelasticity. The color variation shows the different values of Young's modulus.}
  \label{fig:gem2}
\end{figure}
For the second geometry, we introduce $2601$ collocation points to minimize the loss functions (see the left-hand side of Fig.\,\ref{fig:gem2}). The corresponding mesh, which discretizes the domain in order to solve the thermoelasticity problem by means of FEM, is depicted on the right-hand side of Fig.\,\ref{fig:gem2}. To evaluate the response of the system, $10201$ points are employed. In this case, approximately $75\,\%$ of points have different positions concerning the collocation points. The latter addresses the applicability of the NN to interpolate the solution. The material parameters utilized in the second boundary value problem are stated in Tab.\,\ref{tab:par2}.
\begin{table}[H]
	\centering
	\begin{tabular}{ l l  } \hline
	\multirow{1}{*}{}         & Value/Unit    \\ \hline \hline 
	\textbf{Thermal elasticity problem of geometry 2} \\
	 Matrix Young's modulus and Poisson's ratio ($E_{\text{mat}}$, $\nu_{\text{mat}}$)  & ($1.0$ GPa, $0.3$)  \\
 	Inclusion Young's modulus and Poisson's ratio ($E_{\text{inc}}$, $\nu_{\text{inc}}$)  & ($0.5$ GPa, $0.3$)  \\
 	Matrix heat conductivity (${k}_{\text{mat}}$)  & $1.0$~W/mK  \\
   	Inclusion heat conductivity (${k}_{\text{inc}}$)  & $0.5$~W/mK  \\
    \hline
	\end{tabular}
	\caption{Geometry 2 input parameters for the linear thermal elasticity problem}
	\label{tab:par2}
\end{table}
%-------------------------------------
% coupled training geomtry 2
\subsubsection{Coupled training of the second geometry} \label{sec:coupl_2nd_geo}
The loss functions according to the thermoelasticity problem are minimized by the coupled training procedure for the second geometry. The results are reported in Fig.\,\ref{fig:mech_seq_tanh_5_40} where we observe that the PINNs mixed formulation is capable of predicting the response of this multi-physical problem, despite the shape of heterogeneity and also the complex deformation mode of inclusions. However, on the left-hand side of the deformed configuration, one can observe that the left boundary condition is not satisfied completely. The latter is the main source of error for the mechanical field responses of the network. See the first row in Fig.\,\ref{fig:mech_seq_tanh_5_40}, where the left Dirichlet boundary condition is not fully satisfied in certain regions. A remedy for tackling this problem is using hard constraint boundary conditions or adding extra collocation points at boundaries which will be discussed in this work.

Similar to the results of the first geometry, the difference is accumulated mainly at interface regions. The averaged relative errors for the $u_x$ and $\sigma_{x}$ are $1.6\,\%$, and $3.3\,\%$, respectively. Again, we observe the localization of the error in very small regions around inclusions.

For the temperature prediction, the error lives near the left boundary while for the fluxes right boundary is also critical, see Fig.\,\ref{fig:tem_seq_tanh_5_40}. For the mentioned values, the averaged relative errors are $1.3\,\%$,  $1.8\,\%$. 

\color{black}
In Figure~\ref{fig:refine}, we conducted further investigations of the same boundary value problem  by incorporating additional collocation points on the left boundary. We added about 1000 extra points on this edge. This procedure is analogous to mesh refinement in the finite element method. As a result, we were able to reduce the maximum obtained errors by a factor of 1.6 and 3.0 for the displacement and temperature fields, respectively. Moreover, we observed a reduction in the overall average error across the entire domain, indicating the significance of satisfying the boundary conditions.
\color{black}

\begin{figure}[H] 
  \centering
  \includegraphics[width=0.95\linewidth]{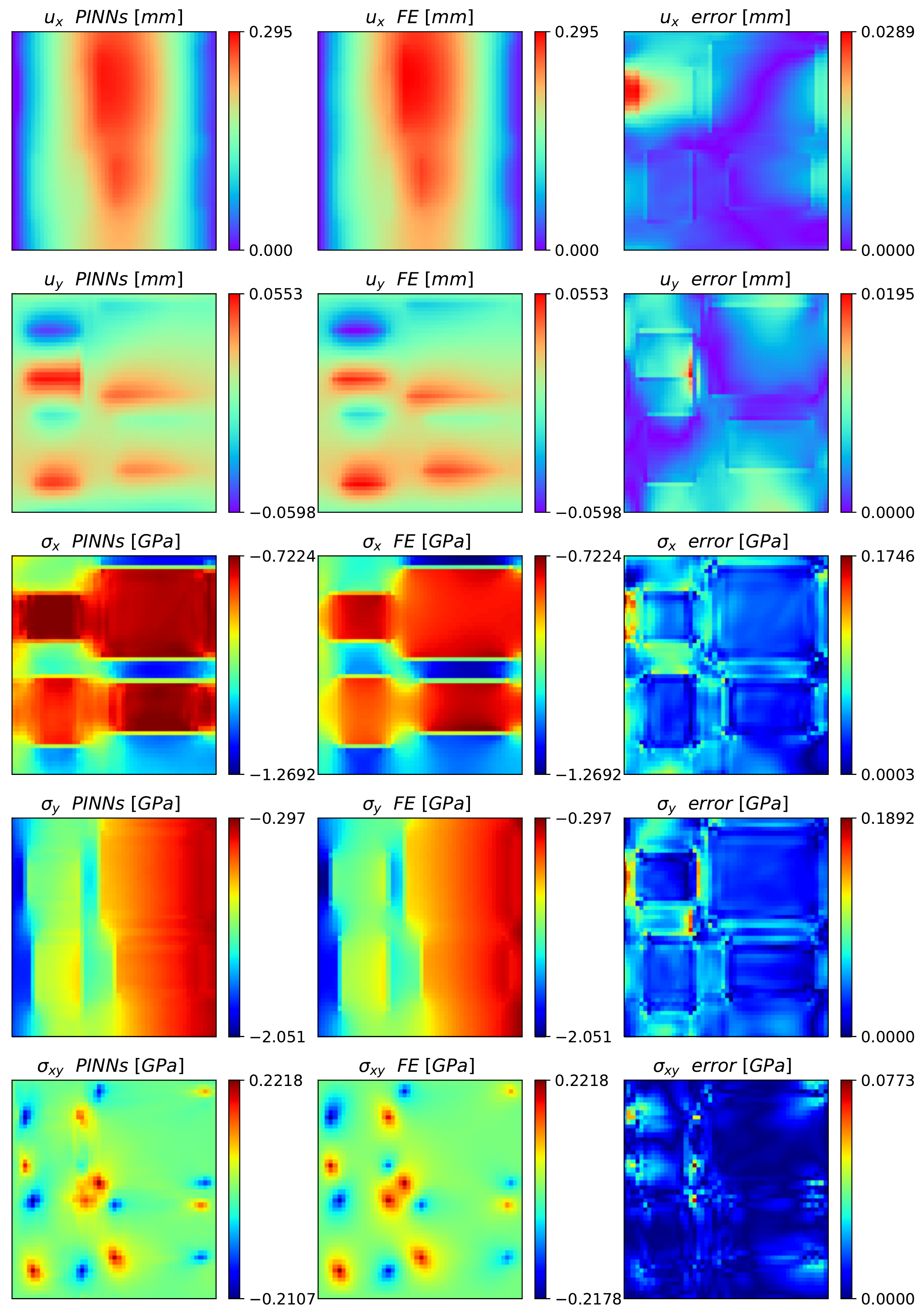}
  \caption{The mechanical field results of the thermoelasticity for geometry 2 using coupled training. Left: PINNs, Middle: FEM, Right: FEM with PINNs difference.}
  \label{fig:mech_seq_tanh_5_40}
\end{figure}

% using coupled training. Left: PINNs, Middle: FEM, Right: FEM with PINNs difference.

\begin{figure}[H] 
  \centering
  \includegraphics[width=0.9\linewidth]{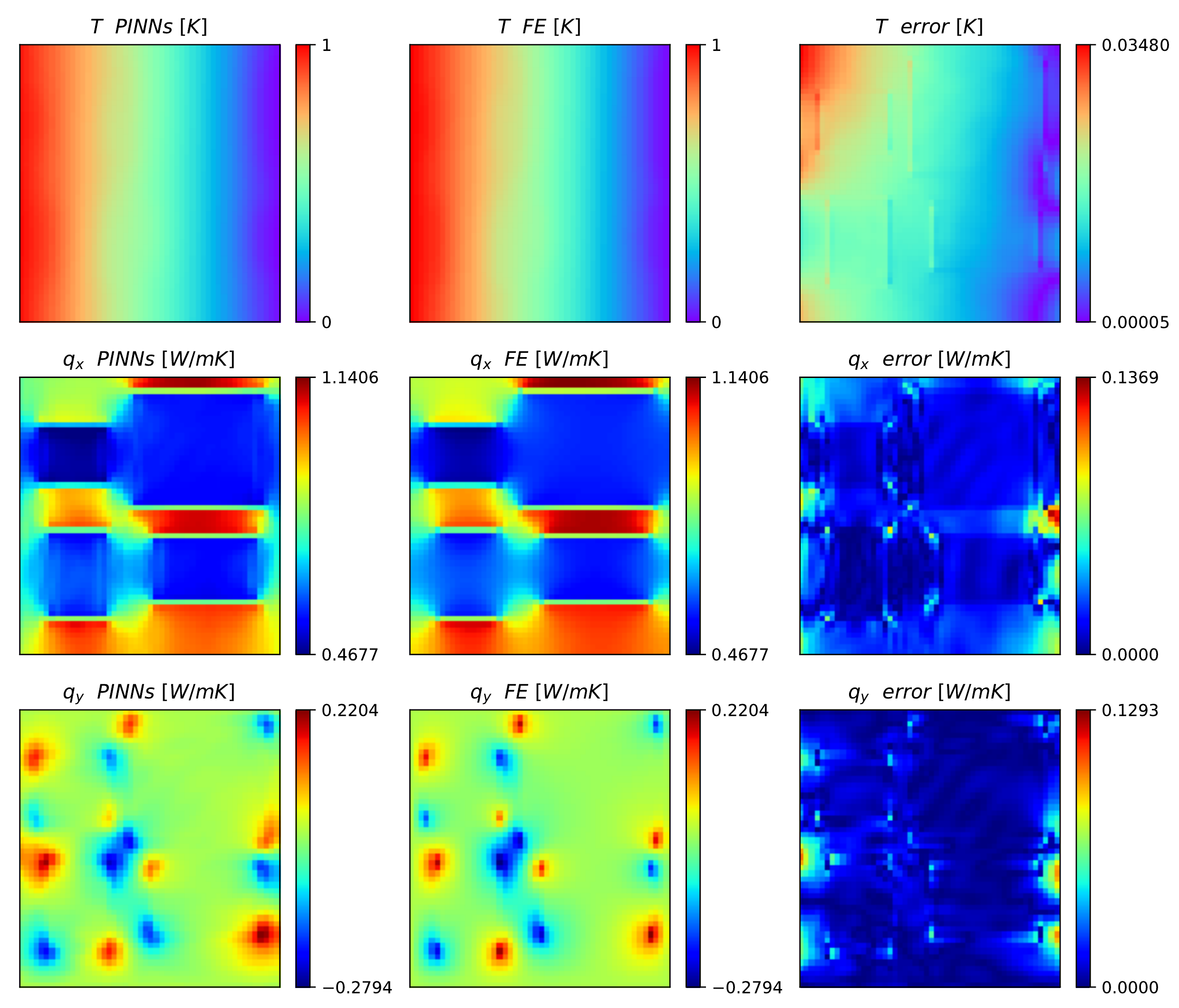}
  \caption{The thermal field results of the thermoelasticity for the second geometry using coupled training. Left: PINNs, Middle: FEM, Right: FEM with PINNs difference.}
  \label{fig:tem_seq_tanh_5_40}
\end{figure}

\begin{figure}[H] 
  \centering
  \includegraphics[width=0.9\linewidth]{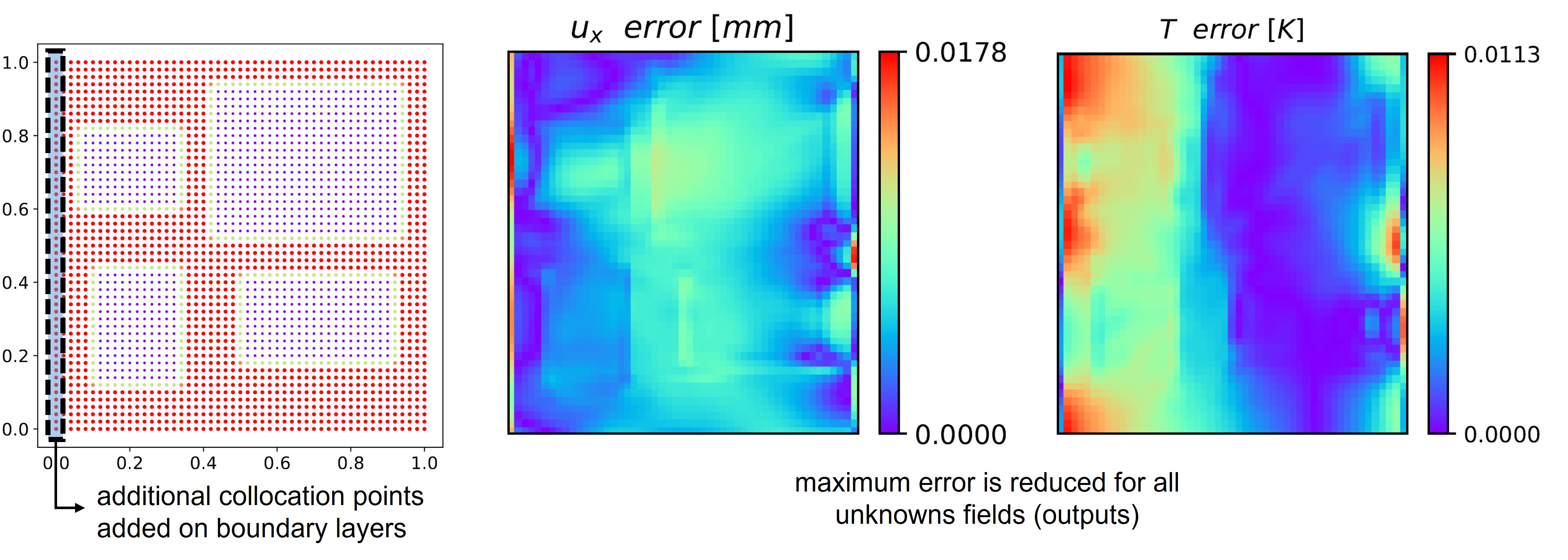}
  \caption{Refinement of collocation points on the left boundary resulted into reduction of error.}
  \label{fig:refine}
\end{figure}

%============================================
% Sequential section geometry 2
%
\subsubsection{Sequential training of the second geometry}
The same number of collocation points as in Sec.\,\ref{sec:coupl_2nd_geo} are utilized to minimize the loss function in the sequential training approach. The corresponding results of the network are partially plotted in Fig.\,\ref{fig:sectiongeo2}.

We also observed that the same issue occurs also for this training strategy and the Dirichlet boundary condition of the left-hand side is not fully satisfied. In general, the results are in good agreement with the results from FEM, the averaged relative difference is about $0.49\,\%$, and $2.12\,\%$ which are for $u_x$ and $\sigma_{x}$, respectively. 
For the thermal field, relative averaged errors lie at $0.7\,\%$. This value for the heat flux in $y$-direction is $1.82\,\%$.

% The maximum calculated relative errors for the mentioned fields are $2.8\,\%$, and $10.19\,\%$.

For carrying out a closer inspection of the results employing coupled and sequential methods, the components of the mechanical field outputs and heat flux in $x$-direction as well as the temperature field are compared by drawing a section at $x=0.5 \text{mm}$. The mentioned sections are made in such a way that at least they go through one inclusion part. The results of PINNs and coupled training are compared by having FE analysis results as the reference solution in Figs.\,\ref{fig:sectiongeo2}. In Appendix A, a mesh convergence analysis was conducted to ensure the accuracy of the reference solution obtained through the FEM. The results of sequential training slightly outperform the coupled training results with respect to the FEM. More specifically, near the inclusion, the results of the sequential training are closer to the reference solution. The related relative difference values mimic the outperformance of the sequential training strategy.

\begin{figure}[H] 
  \centering
  \includegraphics[width=0.99\linewidth]{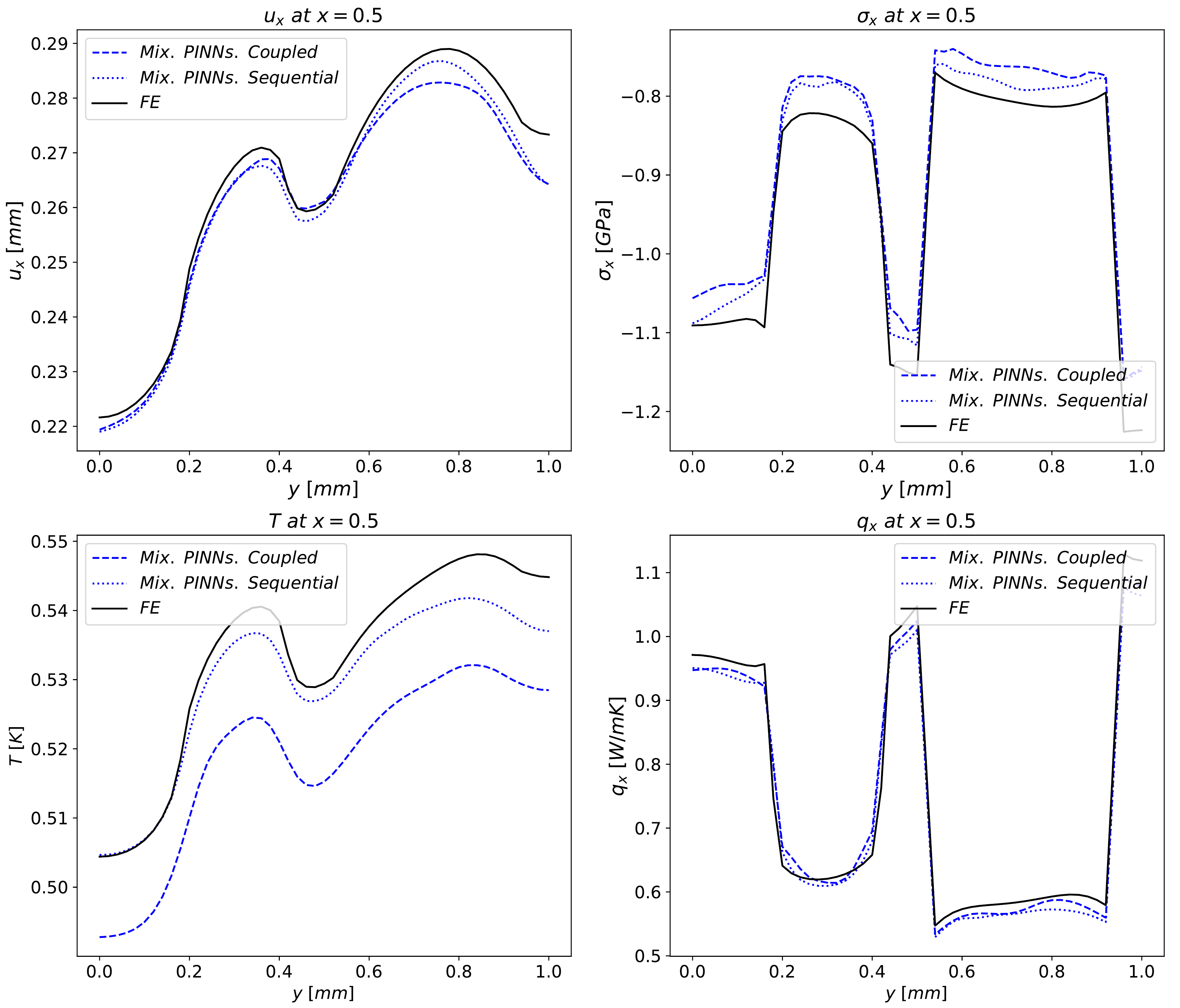}
  \caption{Comparison of the network's outputs through the section at $x=0.5 \text{mm}$ of the geometry 2.}
  \label{fig:sectiongeo2}
\end{figure}

%%%%%%%%%%%%%%%%%%%%%%%%%%%%%%%%%

\subsection{A comprehensive analysis for improved model performance}
In this section, we discuss the concept of hard boundary conditions, which involves adjusting the network's outputs to fulfill the prescribed boundary conditions. This process effectively eliminates prediction errors near fixed boundaries for both displacements and temperature fields, as demonstrated in \cite{lu2021hc} and \cite{alkhadhr2023}. Additionally, the impact of the network's hyperparameters and choice of optimizer on the accuracy and efficiency of the trained network in solving thermomechanical problems using the proposed mixed PINNs formulation is addressed.

To ensure that the satisfaction of the Dirichlet boundary conditions for the second geometry, the outputs of the network are modified as
\begin{align}
\label{hardboundary} %%%%%%%%%%%%%%%
\hat{u}_x&=u_x(x-1)(x), \\
\hat{u}_y&=u_y(y-1)(y), \\
\hat{T}_y&=T(x-1)(x)+1-x.
\end{align}
In Eq.\,(\ref{hardboundary}), the notation $\left(\hat{\bullet}\right)$ represents how the neural network outputs $\left({\bullet}\right)$ are adjusted to ensure the trained network satisfies the Dirichlet boundary conditions. The training process involves minimizing other loss functions through coupled training. To compare the efficiency of the method, training is performed using two different optimizers: Adam optimizer and L-BFGS, which is based on the quasi-Newton method.
The training is conducted using 2601 collocation points, as depicted in Fig.\,\ref{fig:gem2}.
\begin{figure}[H] 
  \centering
  \includegraphics[width=1.02\linewidth]{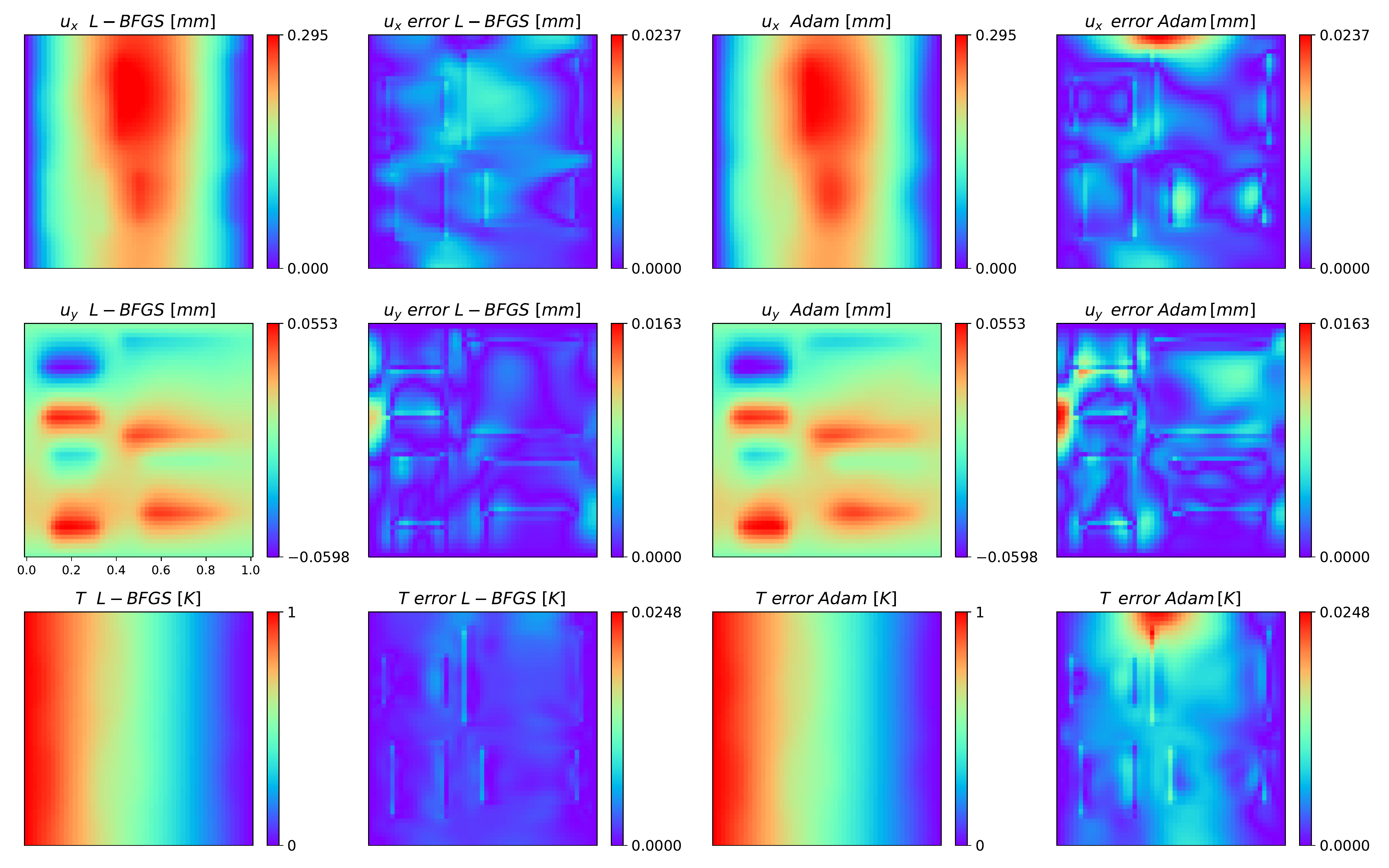}
  \caption{Influence of applying hard Dirichlet boundary conditions on the prediction of displacements and temperature fields by employing L-BFGS and Adam optimizers.}
  \label{fig:hardboundary}
\end{figure}
Fig.\,\ref{fig:hardboundary} illustrates the impact of applying hard boundary constraints on the mechanical and temperature fields, resulting in reduced errors compared to using soft constraint boundary conditions. Moreover, when utilizing the quasi-Newton algorithm optimizer (L-BFGS), the errors decrease significantly to almost one-third of those observed during coupled training with the Adam optimizer. 
Fig.\,\ref{fig:loss_comp}. shows that the total loss yields a lower value for the L-BFGS optimizer compared to Adam, indicating improved performance.
\begin{figure}[H] 
  \centering
  \includegraphics[width=0.99\linewidth]{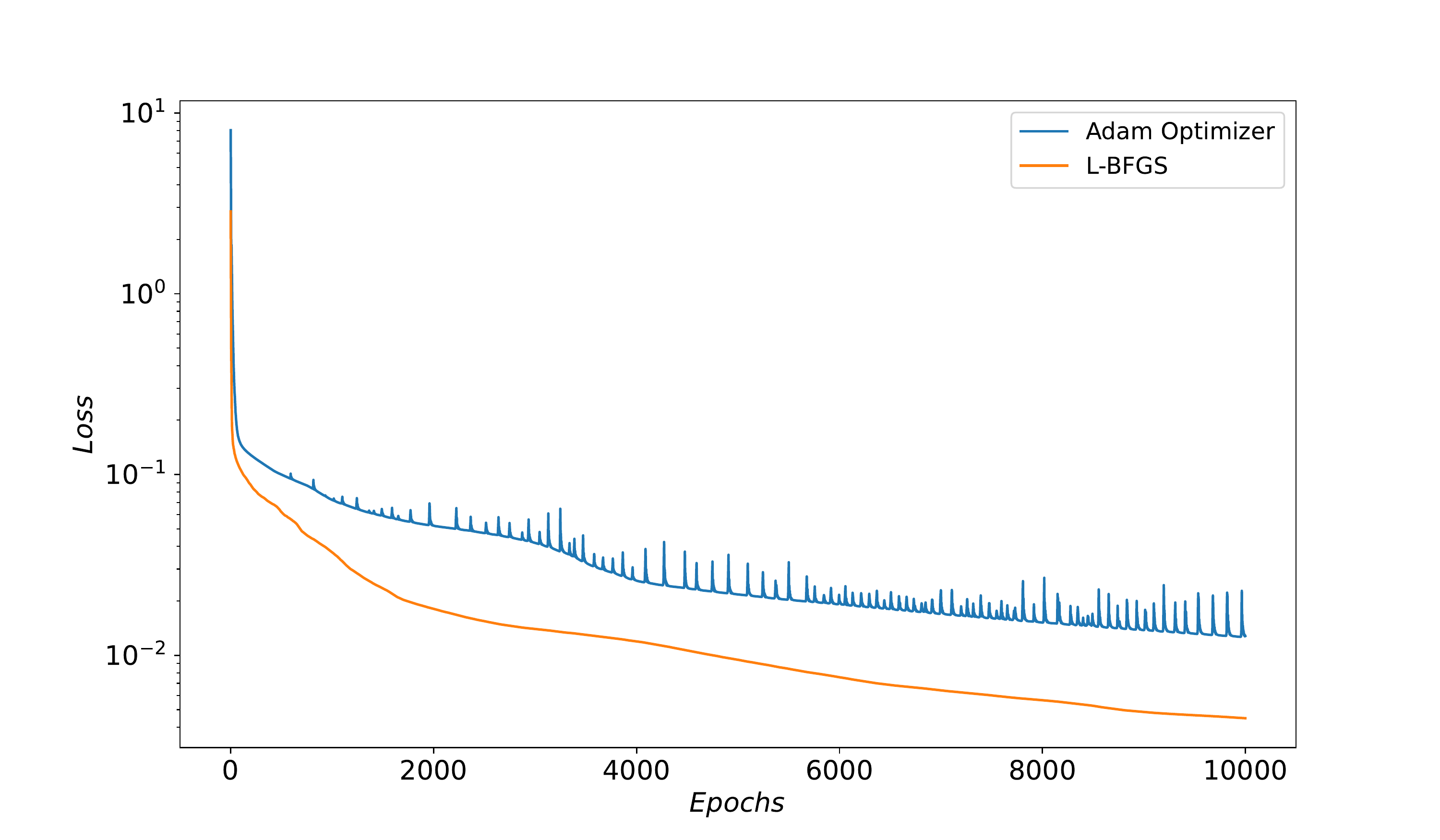}
  \caption{A comparison of total losses between Adam and L-BFGS optimizers for hard constraint network for $10000$ epochs (iterations).}
  \label{fig:loss_comp}
\end{figure}
To investigate the impact of network hyperparameters on the effectiveness and efficiency of the network architecture, we conducted a study on the number of hidden layers and their influence on the network's performance. We utilized five different network architectures in which the number of hidden layers varied from one to six hidden layers. All networks had 40 neurons in each hidden layer, and the layers are fully connected.
The results presented in Fig.\,\ref{fig:layer_study} demonstrate that the network with five layers consistently outperforms the others in terms of various quantities, except for $\sigma_x$, where the error is highly localized, see Fig.\,\ref{fig:error_arch}.

After determining the optimal number of layers for the network, we further investigate the impact of the number of neurons in each layer. Following a similar approach to the previous study, we keep the number of hidden layers fixed at five and train additional networks from five to sixty neurons per hidden layer. The results are presented in Fig.\,\ref{fig:neuron_study}.

Moreover, the computational time of each epoch (iteration) for each network architecture is recorded in Tab.\,\ref{tab:computational_time}.  The reported time is the average time taken for one hundred epochs (iterations).

\begin{figure}[H] 
  \centering
  \includegraphics[width=1.01\linewidth]{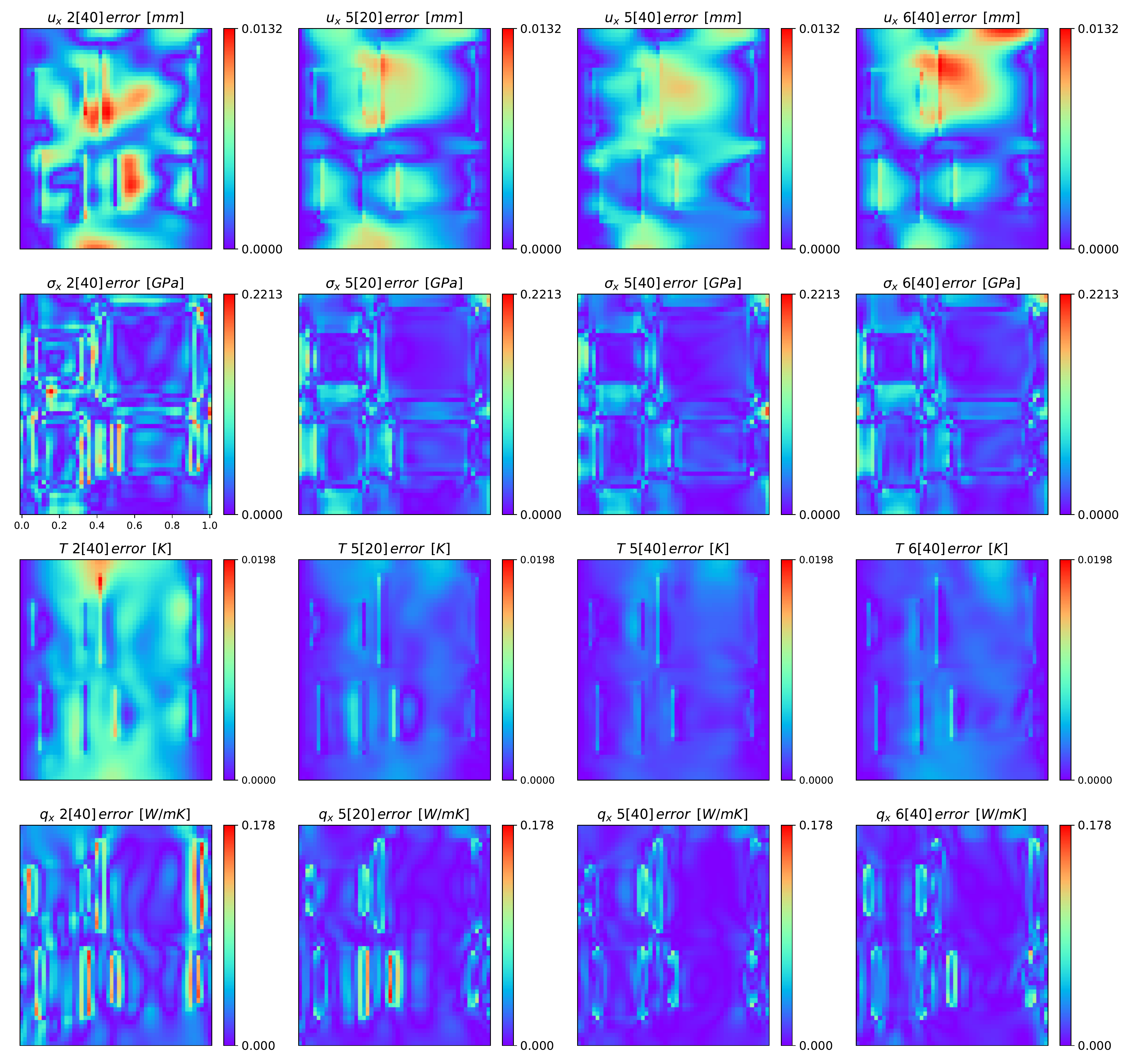}
  \caption{A comparison of different architecture for quantities of interest, all of them with the same activation function ($\tanh$) and the same number of iterations (10000).}
  \label{fig:error_arch}
\end{figure}

\textbf{Remark 7} The boundary value problem for geometry two requires approximately four hours to train using the Adam optimizer with one million iterations. However, by switching to the L-BFGS optimizer and applying hard constraints, we achieve nearly identical results with just ten thousand iterations, reducing the training time to around forty minutes on the Apple M2 Pro platform.

\begin{figure}[H] 
\centering
\includegraphics[width=0.8\linewidth]{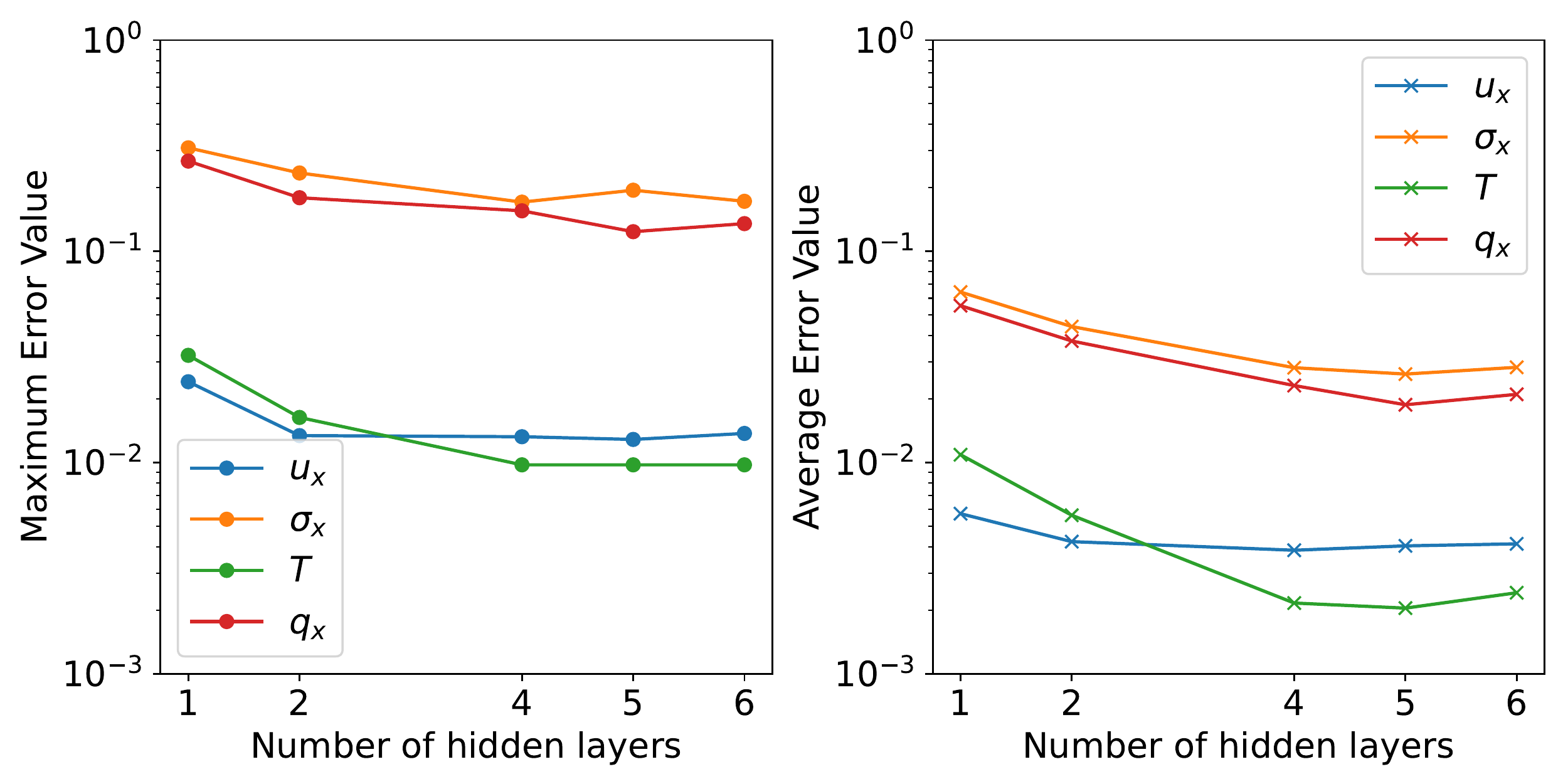}
\caption{Comparison of maximum and average errors of quantities of interest for different numbers of neurons in each hidden layer (with 5 hidden layers, each containing 40 neurons).}
\label{fig:layer_study}
\end{figure}

\begin{figure}[H] 
\centering
\includegraphics[width=0.8\linewidth]{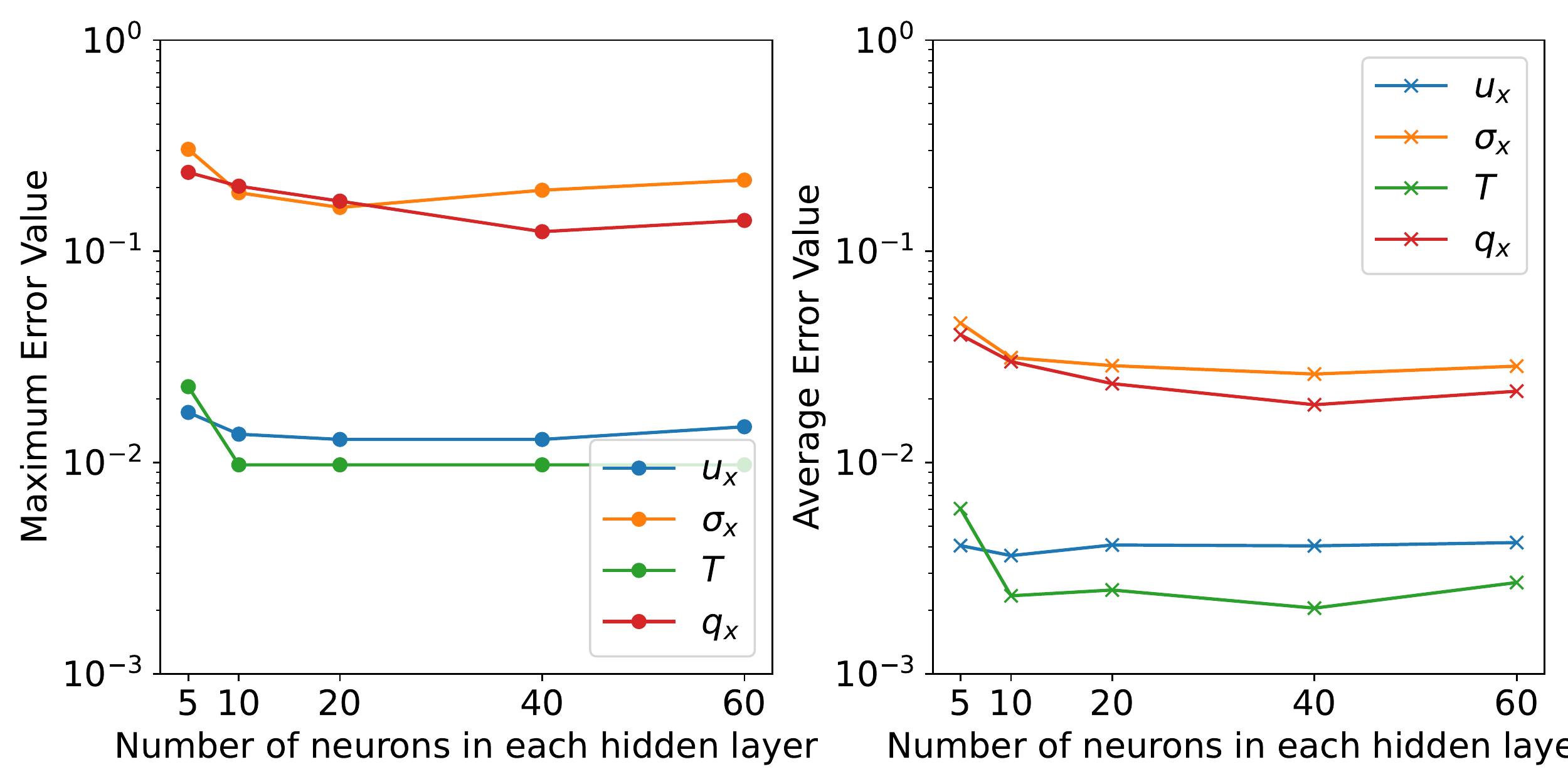}
\caption{Comparison of maximum and average errors of quantities of interest for different numbers of neurons in each hidden layer (number of hidden layers is fixed at $5$).}
\label{fig:neuron_study}
\end{figure}

\begin{table}[htb]
\centering
\begin{tabular}{|c||ccccc|} 
\hline
\diagbox{Layers}{Neurons} & 5 & 10 & 20 & 40 & 60  \\ \hline\hline
1 & 0.2467 & 0.2492      &  0.2526     & 0.2712  & 0.2821 \\
2 & 0.2595      &  0.26685    & 0.2706   & 0.2791 & 0.28845  \\
4 & 0.2936 & 0.2948 & 0.3055 & 0.31865 & 0.3330 \\
5 & 0.3141 & 0.3136 & 0.3294 & 0.3409 & 0.3600 \\
6 & 0.3284 & 0.3346 & 0.3493 & 0.3718 & 0.3974 \\ \hline
\end{tabular} 
\caption{Computational time (in seconds) of every epoch comparison for different network architectures by varying the number of hidden layers and neurons with L-BFGS optimizer on Apple M2 pro.}
\label{tab:computational_time}
\end{table}

%%%%%%%%%%%%%%%%%%%%%%%%%%%%%%%%%
\subsection{Parametric learning by combining physics and data}  \label{sec:data}
In this section, we aim to obtain full-field solutions for unseen cases by combining data and physics, and parametric learning. Parametric learning allows the network to generalize its response to different sets of material parameters. For new sets of material parameters, only the network's response needs to be evaluated, eliminating the need for further training, which takes only a fraction of a second $mS$. The latter is enormously faster than solving the system with conventional solvers such as FE. 

For the given BVP, one can change the topology of the microstructure, boundary conditions as well as material properties. For the sake of simplicity, we focus only on different material properties for different involved phases. Therefore, geometry 2 from the previous section is taken with the set of different material parameters. To this end, the ratio of Young's modulus between inclusion and the matrix as well as their thermal conductivity ratio is changing in the range of $1$ to $10$. For each case, the thermoelasticity problem is solved by means of FE analysis, and the corresponding results, later on, are used to train the neural network. The FE data is utilized to facilitate the process of training and reduce computational costs.

For training the network, one needs to insert Young's modulus and thermal conductivity of each point alongside the collocation points' coordinates as inputs for parametric learning. The designed network's architecture is depicted in Fig.\,\ref{fig:ml_arch_TL}.
In addition to the loss based on data, the losses based on the mixed PINNs formulation are taken into account, $\mathcal{L}_T$ and $\mathcal{L}_M$ see Eqs.\,(\ref{eq:Seq_loss_phys1}) and (\ref{eq:Seq_loss_phys2}). The total loss function in the presence of data from FE analysis is written as
\begin{align}
\label{eq:Totalloss_tl}
\mathcal{L}_{total} &= \underbrace{\mathcal{L}_{data}}_{\text{FE simulations}} +~w \left( \underbrace{\mathcal{L}_{T}  + \mathcal{L}_M}_{\text{Physics}}\right).
\end{align}
In Eq.\,(\ref{eq:Totalloss_tl}), $w$ denotes the weight of the physics in the total loss function. The network's predictions are compared for four different cases in which the $w$ parameter is varying from $0.0$ to $1.0$. In the case of $w=0.0$, the data-driven case is investigated. After studying a set of different values for the weight value, the value of $w=0.1$ is taken that reads the best results. 
The training is done by employing the transfer learning procedure where the $\mathcal{L}_{total}$ in Eq.~(\ref{eq:Totalloss_tl}) is minimized considering a specific ratio for Young's modulus $E(x,y)$ and thermal conductivity $k(x,y)$ of different phases. Next, the network's optimum parameters are found for the first set of values after $10\,\text{k}$ epochs. By having the previously tuned parameters, the minimization of the losses is done for the new value of the ratio for $10\,\text{k}$ iterations. 

\begin{figure}[H] 
  \centering
  \includegraphics[width=1.0\linewidth]{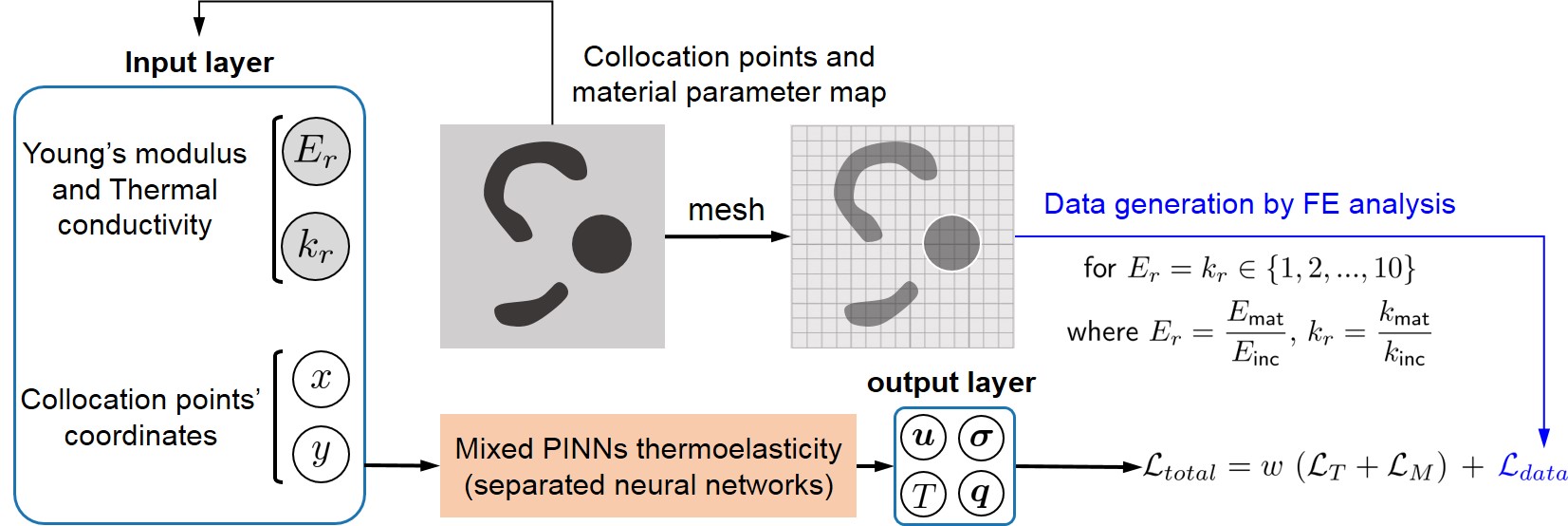}
  \caption{Network inputs and loss functions for the combination of data and physics and utilizing a parametric algorithm.
}
  \label{fig:ml_arch_TL}
\end{figure}

Trained networks based on a combination of data and physics and one based on a purely data-driven approach are then called to predict the system's response for the case of $E_r=K_r=15$. This case can be considered as an extrapolated value for these ratios and it is not in the given training set. 
  
\begin{figure}[H] 
  \centering
  \includegraphics[width=0.95\linewidth]{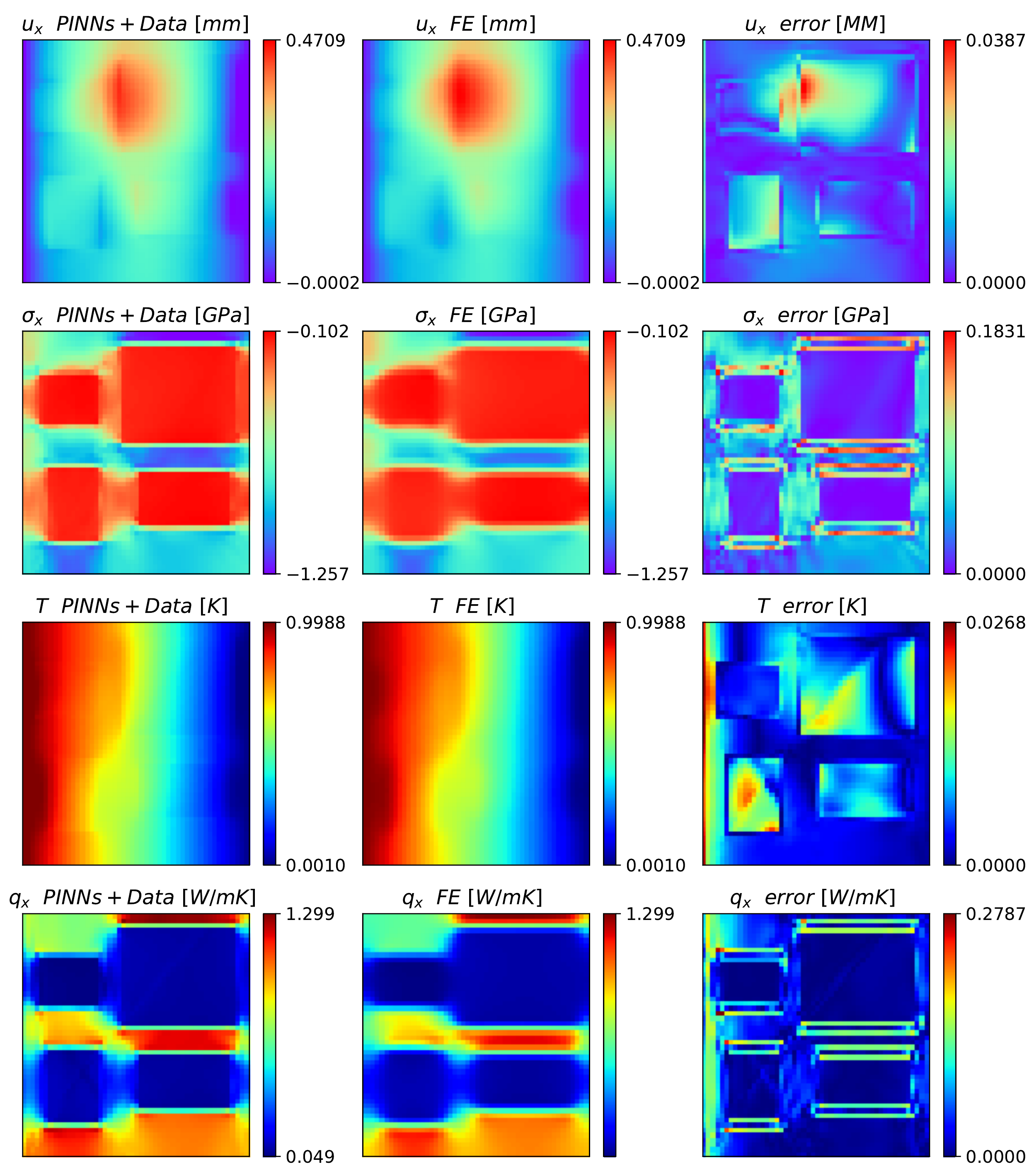}
  \caption{Mechanical and thermal fields results of the thermoelasticity for the case of $E_r=K_r=15$. Left: PINNs + Data ($w=0.1$), Middle: FEM, Right: FEM with PINNs difference.}
  \label{fig:TL_mixed_pinn}
\end{figure}

Figs.\,\ref{fig:TL_mixed_pinn} and~\ref{fig:TL_data} depict the networks' prediction for the preeminent fields for data combined with physics and pure data-driven case, respectively. Figures include $u_x$ and $\sigma_x$ related to the mechanical field and $T$ and $q_x$ for the thermal one.
The combination of data and physics by employing the transfer learning scheme leads to a better prediction for an unseen case of Young's modulus and thermal conductivity ratios (extrapolation).
 
\begin{figure}[H] 
  \centering
  \includegraphics[width=0.95\linewidth]{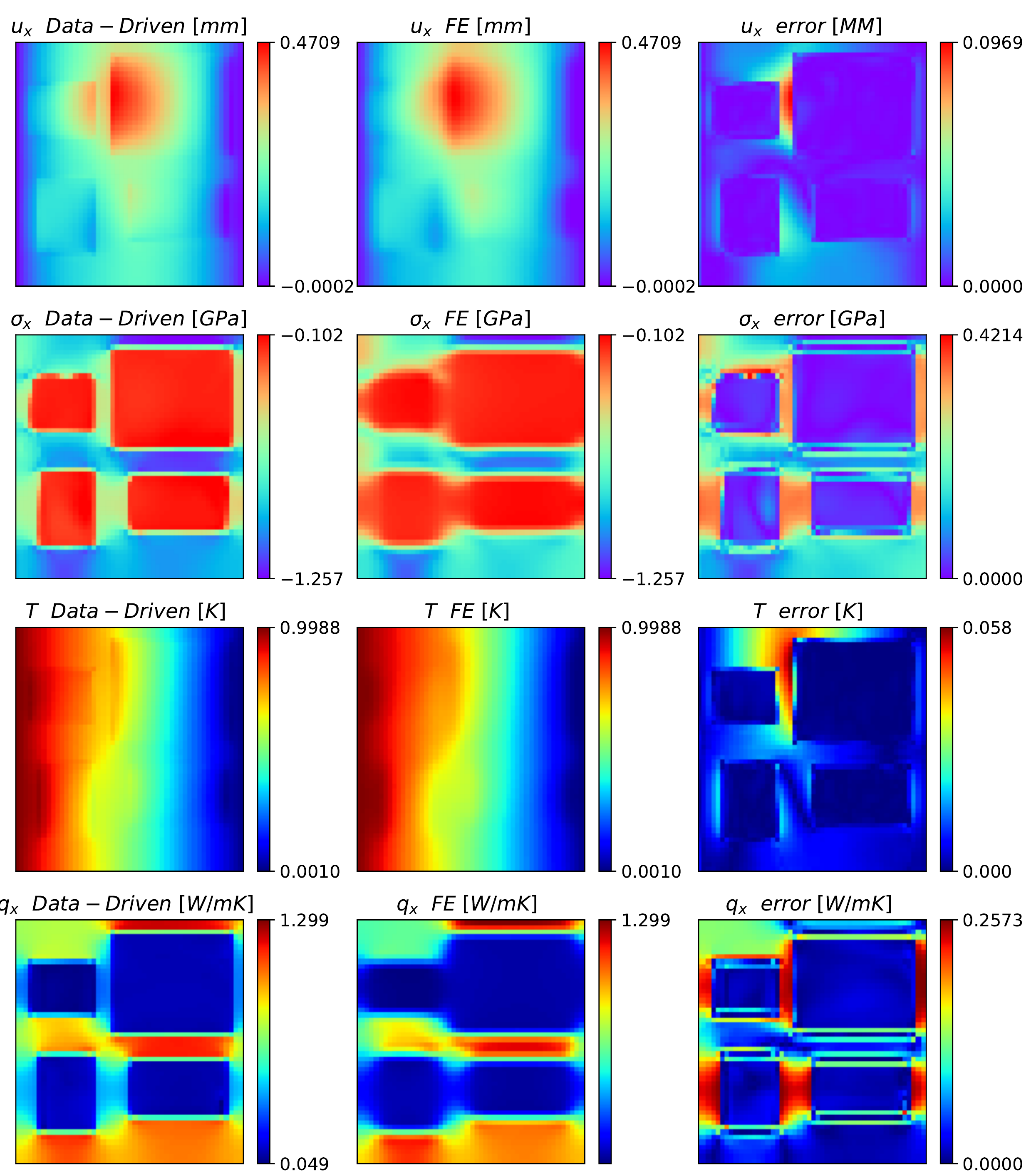}
  \caption{Mechanical and thermal fields results of the thermoelasticity for the case of $E_r=K_r=15$. Left: Data-driven ($w=0.0$), Middle: FEM, Right: FEM with a data-driven difference.}
  \label{fig:TL_data}
\end{figure}

Overall, using physical constraints to enhance the prediction of fields can reduce errors by up to one-third of the error that exists in the pure data-driven case. The maximum error of the heat flux in $x$-direction ($q_x$) is slightly greater when one uses physical constraints, but this error only exists in a few points localized in the domain. Moreover, according to Fig.\,\ref{fig:TL_data}, for the data-driven case, the network's predictions for areas around inclusions are less accurate. 

To have a better comparison, a section is made at $x=0.35~ \text{mm}$. The results from the combination of data and physics as well as the case with the pure data-driven case are reported. The interest outputs in these sections are $u_x$, $\sigma_x$ for the mechanical field. From the thermal field, $T$, and $q_x$ are chosen for comparison.

In general, corresponding results from the mentioned approaches are in good agreement with the reference solution which is computed by the FE analysis, see Fig.\,\ref{fig:sectiongeo1_TL}. The boundary values are approximated better by the mixed PINNs formulation combined with data. Both methods are able to observe the effects of inclusions and the use of physical constraints enhances the accuracy of predictions in areas where heterogeneity exists.

\begin{figure}[H] 
  \centering
  \includegraphics[width=0.9\linewidth]{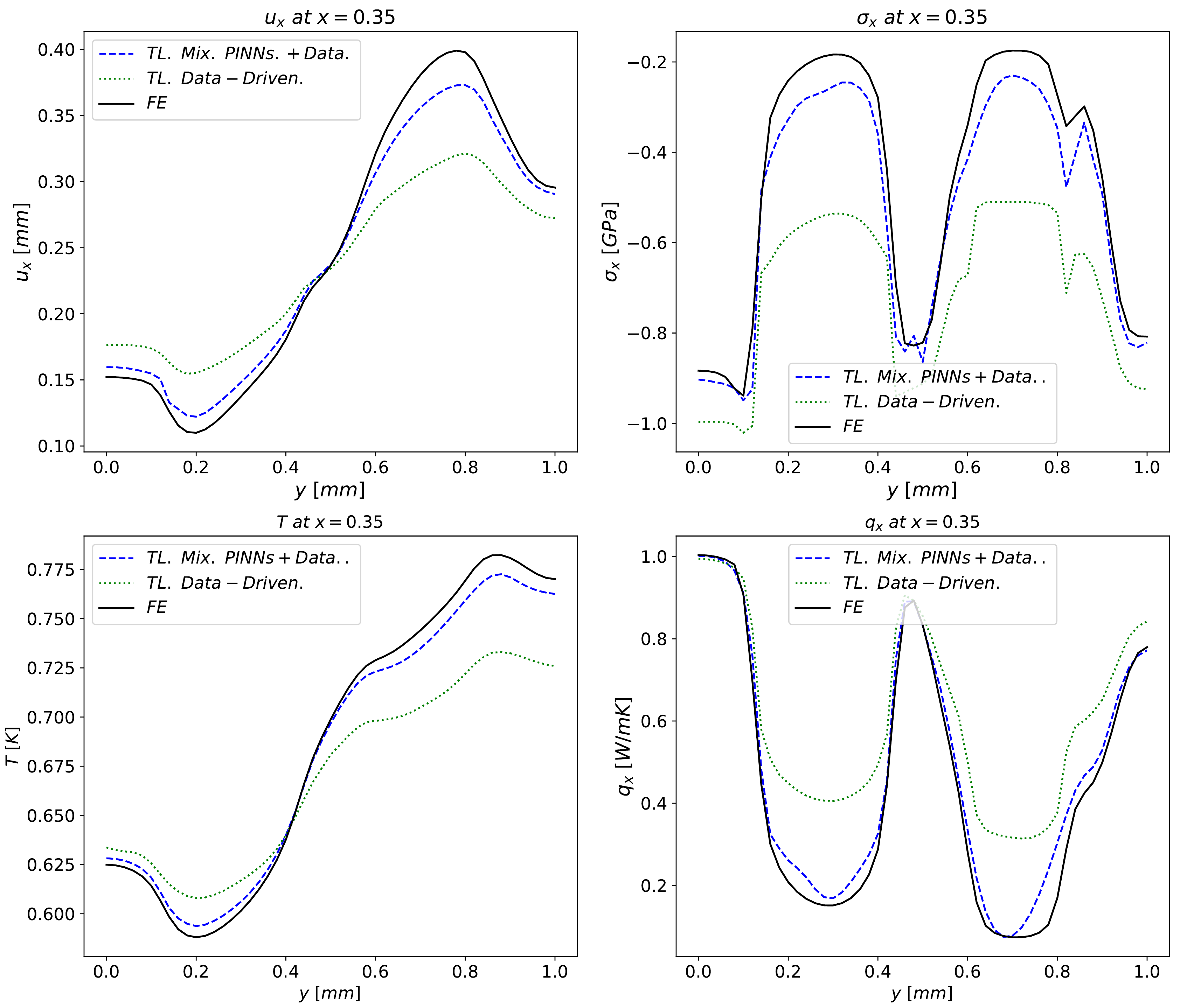}
  \caption{Comparison of main outputs of the networks through the section at $x=0.5 \text{mm}$ of the second geometry using mixed PINNs formulation combined with data utilizing transfer learning training procedure or pure data-driven scheme.}
  \label{fig:sectiongeo1_TL}
\end{figure}

\textbf{Remark 8} The computational cost of the data-driven network employing transfer learning is almost $82\,\%$ less than the network which is trained by combining data and physics. For constructing physical constraints, one needs to differentiate the output parameters with respect to inputs which is the main reason for the higher computational cost. 
\\

%of training since those derivatives should be evaluated in each epoch for every collocation point.
%\newpage
%%%%%%%%%%%%%%%%%%%%%%%%%%%%%%%%%%%%%%%%
%%%%%%%%%%%%%%%%%%%%%%%%%%%%%%%%%%%%%%%%
\section{Conclusion and outlooks}
\color{black}
Current studies demonstrate the effectiveness of mixed PINNs for solving multi-physics problems in heterogeneous domains without any ground truth data, i.e., through unsupervised learning. The mixed PINNs formulation \cite{REZAEI2022PINN, Henkes2022, FUHG2022110839}, which employs only first-order derivatives and a combination of strong and weak forms of equations in the loss functions, enables a more accurate approximation of boundary value problems compared to traditional field approaches that rely solely on the strong form \cite{RAISSI2019} or deep energy methods \cite{SAMANIEGO2020112790} that use only the weak form. Additionally, the study explores sequential training, where loss function minimization is carried out step-by-step by freezing one field at a time (see also \cite{Amini2022}). This method is found to be advantageous in terms of computational efficiency and network accuracy. To validate the network's capabilities, it is utilized to solve a quasi-static thermoelasticity problem, and the results are compared to those obtained via coupled training, where all field losses are minimized simultaneously.

In the subsequent stage, the prescribed network architecture is utilized in conjunction with data to predict outcomes for previously unseen cases. The objective is to enable instant predictions of the system response for varying material properties while keeping computational costs significantly lower than established methods like FEM.
To achieve this, the training process involves to include Young's modulus and thermal conductivity ratios as additional input parameters. The network is trained for various ratio values, and the resulting model is used to predict (extrapolate) values for previously unseen ratios. The trained network can be used to predict the solution to a given boundary value problem with virtually zero computational cost. This can be proved useful in digital twin applications, where a quick system solution is required.
However, additional research is necessary to predict the response of arbitrary microstructures under the same boundary and initial conditions.

In future research, it is crucial to extend the methodology to 3D and compare the performance with other higher-order optimizers like the L-BFGS method (see for example \cite{Abueidda2021}). Extension to the case of elasto-plasticity is also essential in many applications \cite{NIU2023105177, rezaei2023learning}. Additionally, one can focus on solving other coupled partial differential equations, such as the phase-field damage model \cite{REZAEI2022108177, GOSWAMI20TPF}. Moreover, incorporating more complex physics for predicting cracking in multi-physical environments \cite{Ruan2022, REZAEI2023103758} is also proven beneficial.
The trained networks can be leveraged for structural optimizations and optimal design of materials created through additive manufacturing. Additionally, the transfer learning methodology discussed in this paper can be employed to generalize the network for different problem configurations. It would be valuable to combine and compare these ideas with those from operator learning approaches \cite{lu2021learning, Wang2021}. \color{black}
\\ \\
\textbf{Acknowledgements}:\\
Financial support of Subproject A6 of the Trans-regional Collaborative Research Center SFB/TRR 87 as project number 138690629, and
SFB/TRR 339 as the project 453596084 both funded by the German Research Foundation
(DFG) is gratefully acknowledged. The authors also acknowledge financial support by theArbeitsgemeinschaft industrieller Forschungsvereinigungen "Otto von Guericke" e.V.(AiF)through the project grant IGF 21348 N/3.
\\ 
\textbf{Author Statement}:\\
Ali Harandi: Methodology, Software, Writing - Review \& Editing.
Ahmad Moidendin: Methodology, Software, Writing - Review \& Editing.
Michael Kaliske: Supervision, Review \& Editing.
%Bai-Xiang Xu: Supervision, Review \& Editing.
Stefanie Reese: Funding acquisition, Review \& Editing.
Shahed Rezaei: Conceptualization, Methodology, Supervision, Writing - Review \& Editing.

\color{black}
\section*{Appendix A: Convergence study for the finite element calculations}
\label{AppendixA}
To verify that the reference results used in Section \ref{sec:comp_micro} have achieved convergence, a study was conducted by discretizing the domain with three different mesh densities. The material system with multiple inclusions (Geometry 2) was meshed with 50, 100, and 200 elements along each edge. As shown in Fig.~\ref{fig:fe_conv}, the results are almost converged when using 100 elements in each direction. Deviations in the thermal field are negligible and are solely due to the plot scale. 
\begin{figure}[H] 
  \centering
  \includegraphics[width=0.9\linewidth]{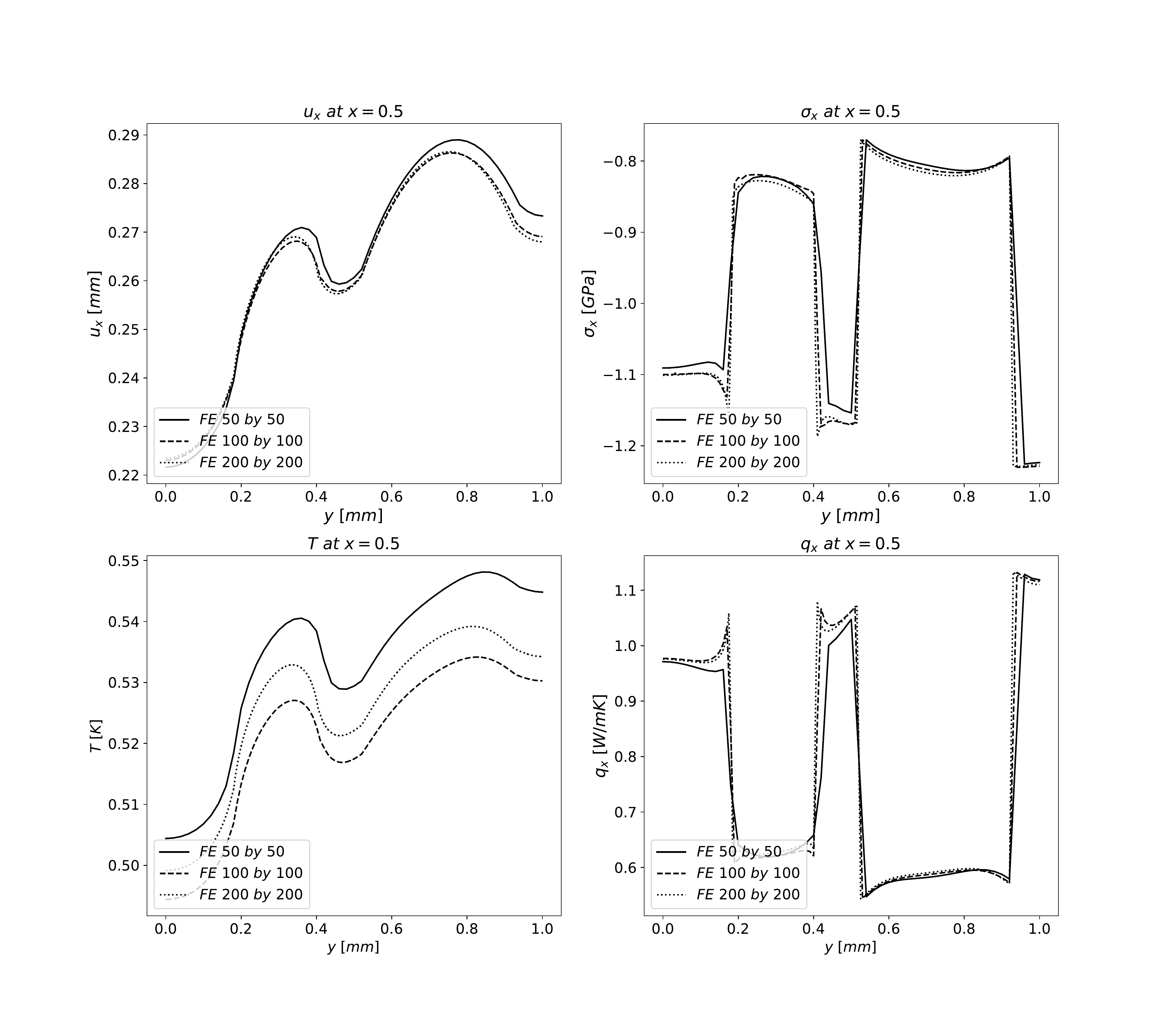}
  \caption{Convergence of the FE results with respect to the number of elements for the second geometry with multiple inclusions.}
  \label{fig:fe_conv}
\end{figure}

%%%%%%%%%%%%%%%%%%%%%%%%%%%%%%%%%%%%%%%
\newpage
\section*{Appendix B: Full field solution by utilizing Dirichlet hard boundary conditions and L-BFGS optimizer}
In this appendix, we present the full field predictions of both mechanical and thermal fields. 
\begin{figure}[H] 
  \centering
  \includegraphics[width=0.80\linewidth]{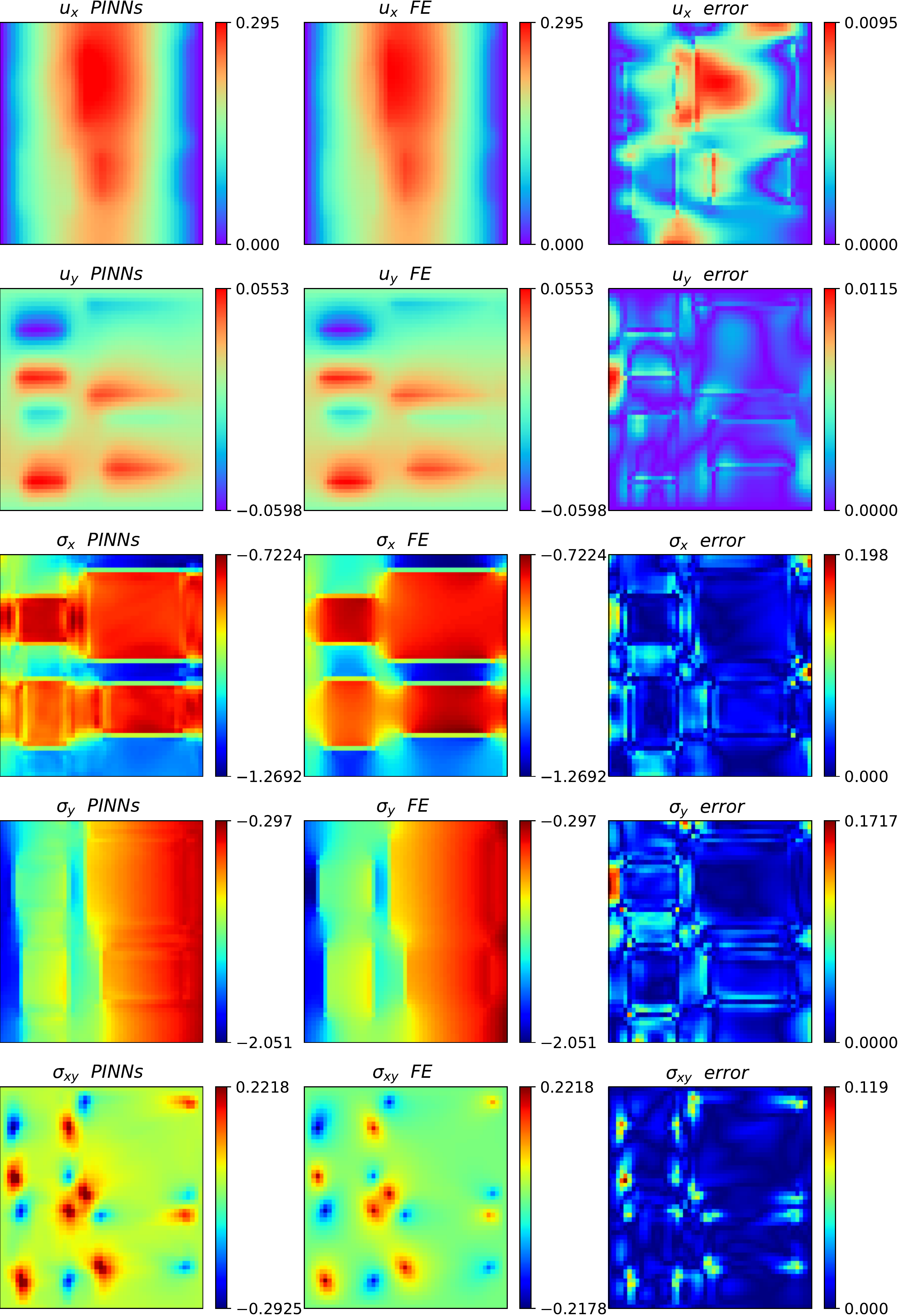}
  \caption{The mechanical field results of the thermoelasticity for geometry 2 using coupled training (L-BFGS optimizer) and hard Dirichlet boundary conditions. Left: PINNs, Middle: FEM, Right: FEM with PINNs difference.}
  \label{fig:mech_hb_tanh_5_40}
\end{figure}
To achieve these predictions, we applied hard boundary constraints and utilized the L-BFGS optimizer in conjunction with the coupled training procedure. The results are showcased after completing 10000 iterations.
By employing hard boundary constraints, we ensure that the predictions are accurate and consistent with Dirichlet boundary conditions.
% using coupled training. Left: PINNs, Middle: FEM, Right: FEM with PINNs difference.

\begin{figure}[H] 
  \centering
  \includegraphics[width=0.9\linewidth]{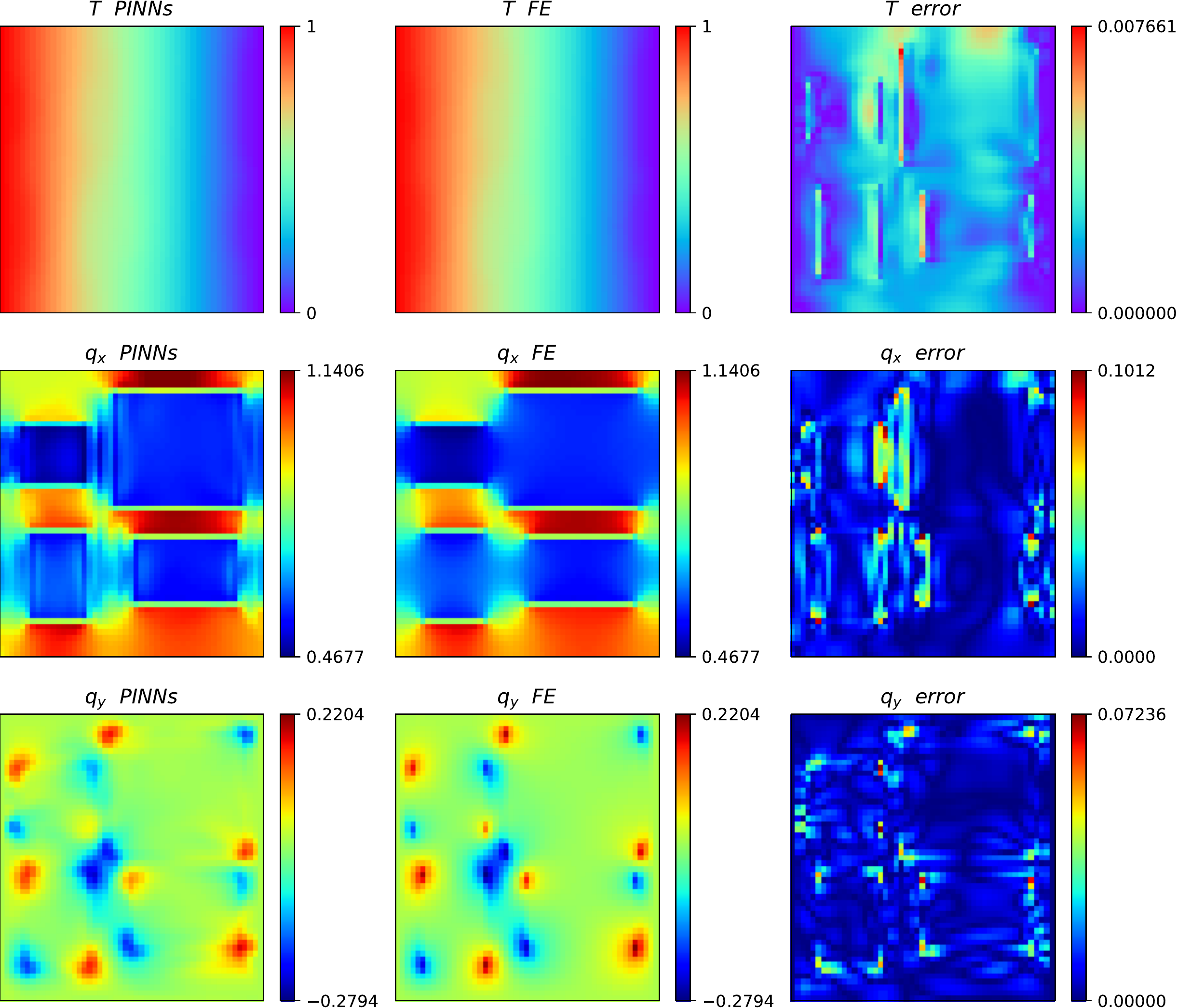}
  \caption{The thermal field results of the thermoelasticity for the second geometry using coupled training (L-BFGS optimizer) and hard Dirichlet boundary conditions. Left: PINNs, Middle: FEM, Right: FEM with PINNs difference.}
  \label{fig:tem_hb_tanh_5_40}
\end{figure}
\color{black}
%%%%%%%%%%%%%%%%%%%%%%%%%%%%%%%%%%%%%%%%
\newpage
\bibliography{Ref}

\end{document}